\documentclass[12pt]{article}
\usepackage{amssymb,amsmath}
\usepackage{graphicx}

\usepackage{rotate}

\usepackage{subfig}
\usepackage{color}
\allowdisplaybreaks
\def\lsim{\ \rlap{\raise 3pt \hbox{$<$}}{\lower 3pt \hbox{$\sim$}}\ }
\def\gsim{\ \rlap{\raise 3pt \hbox{$>$}}{\lower 3pt \hbox{$\sim$}}\ }
\begin{document}
\begin{titlepage}
\newcommand{\AddrLNF}{
  {\it INFN, Laboratori Nazionali di Frascati,C.P. 13, I00044 
    Frascati, Italy}}
\newcommand{\AddrUdeA}{
  {\it Instituto de F\'\i sica, Universidad de Antioquia,
    A.A.{\it{1226}}, Medell\'\i n, Colombia}}
\begin{center}
  \textbf{{\large Purely Flavored Leptogenesis}}
  \\[10mm]
  D. Aristizabal Sierra$^{a}$, 
  Luis  Alfredo Mu\~noz$^{b}$, 
  Enrico Nardi$^{a,b}$
  \vspace{0.4cm}
  \\
  $^a$\AddrLNF.\vspace{0.4cm}\\
  $^b$\AddrUdeA.\vspace{0.4cm}\\
\end{center}
\begin{abstract}
  We study a model for leptogenesis in which the total CP asymmetries in the
  decays and scatterings involving the $SU(2)$ singlet seesaw neutrinos
  $N_\alpha$ vanish ($\epsilon_{N_\alpha}=0$). Leptogenesis is possible due to
  non-vanishing CP violating lepton flavor asymmetries, realizing a situation
  in which the baryon asymmetry is due exclusively to flavor effects.  We
  study the production of a net lepton asymmetry by solving the Boltzmann
  equations specific to this model, and we show that successful leptogenesis
  can be obtained at a scale as low as the TeV. We also discuss constraints on
  the model parameter space arising from current experimental upper limits on
  lepton flavor violating decays.
\end{abstract}
\end{titlepage}
\section{Introduction}
\label{sec:introduction}
Leptogenesis~\cite{Fukugita:1986hr,Luty:1992un} (for a comprehensive
review see ref.~\cite{Davidson:2008bu}) is a theoretical mechanism
that can explain the observed matter-antimatter asymmetry of the
Universe.  An initial lepton asymmetry, generated in the
out-of-equilibrium decays of heavy singlet Majorana neutrinos, is
partially converted in a baryon asymmetry by anomalous sphaleron
interactions~\cite{Kuzmin:1985mm} that are standard model (SM)
processes.  Heavy Majorana singlet neutrinos are also a fundamental
ingredient of the seesaw model~\cite{seesaw}, that provides an elegant
explanation for the suppression of the neutrino masses with respect to
all other SM mass scales.  Leptogenesis can be quantitatively
successful with a neutrino mass scale of the order of the atmospheric
neutrino mass squared difference. This remarkable `coincidence' links
nicely the explanation of neutrino masses and of the baryon asymmetry
within a single framework, and renders the idea that baryogenesis
occurred through leptogenesis a very attractive one.

In the  standard seesaw case, computations of the CP violating (CPV)
asymmetries $\epsilon_{N_\alpha}$ in the decays of the $N_\alpha$
singlet neutrinos include loop diagrams in which Majorana states
appear in the internal lines~\cite{co96}, and thus are lepton number
violating quantities~\cite{Kolb:1979qa}.  This is the reason why
leptogenesis can proceed even when it is assumed that only one lepton
flavor is relevant, as is the case at large temperatures ($T \gsim
10^{12}\,$GeV).  However, at temperatures below $\sim 10^{12}\,$GeV,
lepton flavor dynamics plays an important role in leptogenesis, and
cannot be neglected~\cite{aba06a-aba06b,Nardi:2006fx} (See
\cite{ba00,en03-fu05} for earlier studies of flavor effects in
leptogenesis, and \cite{Davidson:2008bu,review} for recent reviews).
In particular, in ref.~\cite{Nardi:2006fx} it was pointed out that
leptogenesis can occur even when $\epsilon_{N_\alpha}=0$, provided
that the individual flavored CPV asymmetries $\epsilon_{N_\alpha\to
  L_j}$ (with $j = e,\mu,\tau$) are non vanishing.

Of course, $\epsilon_{N_\alpha}=\sum_j \epsilon_{N_\alpha\to L_j}=0$
means that total lepton number is not violated in $N_\alpha$ decays.
The reason why leptogenesis can still occur even in this case can be
understood by analogy with the generation of a baryon asymmetry
$\Delta B=B-\bar B$ from a lepton asymmetry $\Delta L=L-\bar L$ that,
as is well known, does not require any baryon number violating CP
asymmetry.  Baryon number, or more precisely $\Delta B + \Delta L$, is
in fact violated in the plasma by fast sphaleron reactions, with the
result that part of $\Delta L$ is converted in $\Delta B$ yielding a
ratio $\Delta B/\Delta L = -28/51$.

Similarly, at $T \sim M_{N_\alpha}$ various interactions that are
lepton and lepton flavor number violating occur in the plasma, like
for example $ \Phi\ell_j \leftrightarrow N_\alpha \leftrightarrow \bar
\Phi \bar\ell_k$.  Of course, these reactions must be at least
slightly out-of-equilibrium, otherwise they would quickly drive the
individual $\Delta L_j \to 0$.  However, the important point here is
that generically these (washout) reactions proceed with different
rates for different lepton flavors, erasing more efficiently, say,
$\Delta L_\tau$ than $\Delta L_{e,\mu}$.  As a result, at $T\ll
M_{N_1}$ ($N_1$ being the lightest heavy Majorana neutrino), once all
washout processes are switched off, quite generically $\sum_j \Delta
L_j = \Delta L \neq 0$ results.  This scenario, in which leptogenesis
can proceed solely because of flavor effects, is what we call Purely
Flavored Leptogenesis (PFL).  It is worth noticing that PFL realizes
the Sakharov conditions~\cite{Sakharov:1967dj} in a slightly different
way that standard leptogenesis, since violation of lepton number
occurs only in the washouts, while CP is violated only in the flavor
charges, and the two conditions are thus disentangled.

Ref.~\cite{AristizabalSierra:2007ur} analyzed the issue of the interplay
between the lepton number breaking scale and the breaking scale of a flavor
symmetry (of the Froggatt-Nielsen type~\cite{Froggatt:1978nt}).  It was found
that in the case when the flavor symmetry is still unbroken during the
leptogenesis era, but the vectorlike messengers masses are larger than the
Majorana neutrino mass, the total CP asymmetry vanishes and a PFL scenario
arises. In this paper we show that the PFL model of
ref.~\cite{AristizabalSierra:2007ur} can indeed succeed in producing the cosmological
baryon asymmetry. Interestingly, in this model there is an upper limit on the
leptogenesis temperature fixed by the requirement that leptogenesis must occur
in the flavored regime ($T\lsim 10^{12}\,$GeV) but, differently from the
standard case, there is no lower limit and, as we will show, leptogenesis can
be successful at a scale as low as the TeV.

The rest of the paper is organized as follows: in section~\ref{sec:model} we
recall the main features of the model, we give the expressions for the flavor
CPV asymmetries and we discuss an important rescaling property of the CPV
asymmetries that leaves unaffected the washout rates.  In section \ref{sec:BE}
we write down the Boltzmann Equations (BE) for the model and we present the
main results.  In section \ref{sec:lfv-processes} we study some relations
between the flavor violating parameters of the model and the low energy limits
on lepton flavor violating processes.

In our model washout and asymmetries in decays and scatterings occur at the
same order in the couplings, and thus the derivation of the BE differs from
the standard case in a non-trivial way.  In appendix \ref{sec:appendixA} we
present a detailed derivation of the BE, that relies on the formalism
introduced in ref.~\cite{Nardi:2007jp} to deal in a proper way with CPV
asymmetries in scatterings.  In appendix \ref{sec:appendixB} we collect some
definitions and useful formulae.

\section{The Model}
\label{sec:model}
The model we consider here \cite{AristizabalSierra:2007ur} is a simple
extension of the SM containing a set of $SU(2)_L\times U(1)_Y$
fermion singlets, namely three right-handed neutrinos ($N_\alpha = N_{\alpha
  R} + N_{\alpha R}^c$) and three heavy vectorlike fields ($F_a=F_{aL} +
F_{aR}$). In addition, we assume that at some high energy scale, taken to be
of the order of the leptogenesis scale $M_{N_1}$, an exact $U(1)_X$ horizontal
symmetry forbids direct couplings of the lepton $\ell_i$ and Higgs $\Phi$
doublets to the heavy Majorana neutrinos $N_\alpha$.  At lower energies,
$U(1)_X$ gets  spontaneously broken by the vacuum expectation value 
$\sigma$ of a $SU(2)_L$ singlet scalar field $S$.  Accordingly, the Yukawa
interactions of the high energy Lagrangian read
\begin{equation}
  \label{eq:lag}
  -{\cal L}_Y = 
  \frac{1}{2}\bar{N}_{\alpha}M_{N_\alpha}N_{\alpha} +
  \bar{F}_{a}M_{F_a}F_{a} +
  h_{ia}\bar{\ell}_{i}P_{R}F_{a}\Phi + 
\bar{N}_{\alpha}
\left(  \lambda_{\alpha a} +   \lambda^{(5)}_{\alpha a}\gamma_5\right) 
     F_{a}S 
  + \mbox{h.c.}   
\end{equation}
We use Greek indices $\alpha,\beta\dots =1,2,3$ to label the heavy Majorana
neutrinos, Latin indices $a,b\dots =1,2,3$ for the vectorlike messengers, and
$i, j, k, \dots$ for the lepton flavors $e,\mu,\tau$.  Following reference
\cite{AristizabalSierra:2007ur} we chose the simple $U(1)_X$ charge
assignments $X(\ell_{L_i},F_{L_a},F_{R_a})=+1$, $X(S)=-1$ and
$X(N_{\alpha},\Phi)=0$.  This assignment is sufficient to enforce the absence
of $\bar N \ell \Phi$ terms, but clearly it does not constitute an attempt to
reproduce the fermion mass pattern, and accordingly we will also avoid
assigning specific charges to the right-handed leptons and quark fields that
have no relevance for our analysis.  The important point is that it is likely
that any flavor symmetry (of the Froggatt-Nielsen type) will forbid the the
same tree-level couplings, and will reproduce an overall model structure
similar to the one we are assuming here. Therefore we believe that our
results, that are focused on a new realization of the leptogenesis mechanism,
can hint to a general possibility that could well occur also in a complete
model of flavor.

As it was discussed in~\cite{AristizabalSierra:2007ur}, depending on the
hierarchy between the relevant scales of the model
($M_{N_1},\,M_{F_a},\,\sigma$), quite different {\it scenarios} for
leptogenesis can arise.  PFL arises when the relevant scales satisfy the
hierarchy $\sigma < M_{N_1}< M_{F_a}$ that is, when the flavor symmetry
$U(1)_X$ is still unbroken during the leptogenesis era and at the same time
the messengers $F_a$ are too heavy to be produced in $N_1$ decays and
scatterings, and can be integrated away.  As is explicitely shown by the last
term in eq.~(\ref{eq:lag}), in general the vectorlike fields can couple to the
heavy singlet neutrinos via scalar and pseudoscalar couplings, In
ref.~\cite{AristizabalSierra:2007ur} it was assumed for simplicity a strong
hierarchy $\lambda\gg \lambda^{(5)}$ which allowed us to neglect all the
$\lambda^{(5)}$.  However, in all the relevant quantities (scatterings, CP
asymmetries, light neutrino masses) at  leading order the scalar and
pseudoscalar couplings always appear in the combination $\lambda +
\lambda^{(5)}$, and thus such an assumption is not necessary.  The replacement
$\lambda \to \lambda + \lambda^{(5)}$ would suffice to include in the analysis
the effects of both type of interactions.

\subsection{Extended seesaw and light neutrino masses}
\label{sec:nmgen}
\begin{figure}[t]
  \centering
  \includegraphics[width=8.5cm,height=3cm]{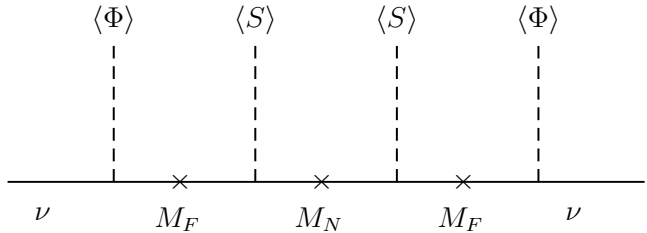}
  \caption{Effective seesaw operator for the light neutrino masses}
  \label{fig:neutrino-massmatrix}
\end{figure}
After $U(1)_X$ and electroweak symmetry breaking the  Lagrangian
eq.~(\ref{eq:lag}) generates masses for the light neutrinos through the
effective mass operator depicted in figure~\ref{fig:neutrino-massmatrix}. The
resulting mass matrix reads~\cite{AristizabalSierra:2007ur}
\begin{equation}
  \label{eq:nmm}
  -{\cal M}_{ij}=
  \left[
    h^{*}\frac{\sigma}{M_{F}}\lambda^{T}\frac{v^{2}}
    {M_{N}}\lambda\frac{\sigma}{M_{F}}h^{\dagger}
  \right]_{ij}
  = \left[
    \tilde{\lambda}^{T}\frac{v^{2}}{M_{N}}\tilde{\lambda}
  \right]_{ij} \,,
\end{equation}
where we have introduced effective seesaw-like couplings  defined as 
\begin{equation}
  \label{eq:seesaw-couplings}
  \tilde{\lambda}_{\alpha i} = 
  \left(
    \lambda \frac{\sigma}{M_F}h^\dagger
  \right)_{\alpha i}=\frac{\sigma}{M_{N_1}}
\left(
    \lambda.r.h^\dagger
  \right)_{\alpha i}\,.
\end{equation}
Note that, differently from standard seesaw, the neutrino mass matrix is of
fourth order in the {\it fundamental} couplings ($h$ and $\lambda$) and
includes an additional suppression factor of $(\sigma/M_F)^2$.

\subsection{$N_1$ decays and CPV asymmetries}
\label{sec:CP}
\begin{figure}[t]
\begin{center}
\includegraphics[width=10cm,height=3cm]{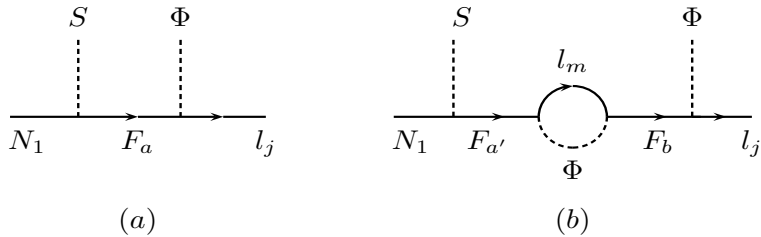}
\end{center}
\caption{Feynman diagrams responsible for the CPV asymmetry.}
\label{fig:fig0}
\end{figure}

Differently from standard leptogenesis in the present case, since $M_F >
M_{N_1}$, two-body $N_1$ decays are kinematically forbidden.  However, via
off-shell exchange of the heavy $F_a$ fields, $N_1$ can decay to the three
body final states $S\Phi l$ and $\bar S\bar\Phi \bar l$. The corresponding
Feynman diagram is depicted in figure \ref{fig:fig0}$(a)$.  At leading order
in $r_a=M_{N_1}/M_{F_a}$, the total decay width reads
\cite{AristizabalSierra:2007ur}
\begin{equation}
  \label{eq:total-decay-width}
 \Gamma_{N_1}\equiv  
\sum_j \Gamma(N_1\to S\Phi l_j + \bar S\bar\Phi \bar l_j) =\frac{M_{N_1}}{192\pi^3}
  \left(
    \frac{M_{N_1}}{\sigma}  
  \right)^2
  (\tilde{\lambda}\tilde{\lambda}^\dagger)_{11}\,.
\end{equation}

As usual, CPV asymmetries in $N_1$ decays arise from the interference between
tree-level and one-loop amplitudes.  As was noted
in~\cite{AristizabalSierra:2007ur}, in this model at one-loop there are no
contributions from vertex corrections, and the only contribution to the CPV
asymmetries comes from the self-energy diagram~\ref{fig:fig0}$(b)$.  Summing
over the leptons and vectorlike fields running in the loop, at leading order in
$r_a$ the CPV asymmetry for $N_1$ decays into leptons of flavor $j$ can be
written as
\begin{equation}
  \label{eq:cp-violating-asymm}
 \epsilon_{1j}  \equiv    \epsilon_{N_1\to\ell_j} = 
  \frac{3}{128\pi}
  \frac{\sum_{m} \mbox{Im} 
    \left[
      \left(
        h r^{2} h^{\dagger}
      \right)_{mj}\tilde{\lambda}_{1m}\tilde{\lambda}^{*}_{1j} 
    \right]}{\left(\tilde{\lambda}\tilde{\lambda}^{\dagger}\right)_{11}}\,.
\end{equation}
Note that since the loop correction does not violate lepton number, the total
CPV asymmetry that is obtained by summing over the flavor of the final state
leptons vanishes~\cite{Kolb:1979qa}, that is $\epsilon_{1}\equiv \sum_j
\epsilon_{1j}=0$.  This is the condition that defines PFL; namely there is no
CPV {\it and} lepton number violating asymmetry, and the CPV lepton flavor
asymmetries are the only seed of the Cosmological lepton and baryon
asymmetries.

It is important to note that the effective couplings $\tilde\lambda$ defined
in eq.~(\ref{eq:seesaw-couplings}) are invariant under the reparameterization
\begin{equation}
  \label{eq:couplingtrans}
  \lambda\to \lambda\cdot (rU)^{-1},\quad
  h^\dagger \to (U r)\cdot h^\dagger\,,
\end{equation}
where $U$ is an arbitrary $3\times 3$ non-singular matrix.  Clearly the light
neutrino mass matrix is invariant under this transformation. Moreover, also
the flavor dependent washout processes, that correspond to tree level
amplitudes that are determined, to a good approximation, by the effective
$\tilde \lambda$ couplings, are left essentially unchanged.\footnote{The
  approximation is exact in the limit of pointlike $F$-propagators 
$(s-M^2_F+iM_F\Gamma_F)\to M^2_F$.}
On the contrary,
the flavor CPV asymmetries eq.~(\ref{eq:cp-violating-asymm}), that are
determined by loop amplitudes containing an additional factor of $h r^2
h^\dagger$, get rescaled as $h r^2 h^\dagger\to h (rUr)^\dagger (rUr)
h^\dagger $.  Clearly, this rescaling affects in the same way all the lepton
flavors (as it should be to guarantee that the PFL conditions
$\epsilon_\alpha\equiv \sum\epsilon_{\alpha j}=0$ are not spoiled), and thus
for simplicity we will consider only rescaling by a global scalar factor
$r.U=U.r=\kappa\,I$ (with $I$ the $3\times 3$ identity matrix) that, for our
purposes, is completely equivalent to the more general
transformation~(\ref{eq:couplingtrans}). Thus, while rescaling the Yukawa couplings
through
\begin{equation}
  \label{eq:coupling-rescaling-gen}
  \lambda\to \lambda \,\kappa^{-1},\quad 
  h^\dagger\to\kappa \,h^\dagger\,,
\end{equation}
does not affect neither low energy neutrino physics nor the washout processes,
the CPV asymmetries get rescaled as:
\begin{equation}
  \label{eq:rescaled-CPV-asymm}
  \epsilon_{1j}\to\kappa^2\epsilon_{1j}\,.
\end{equation}
By choosing $\kappa>1$, all the CPV asymmetries get enhanced as $\kappa^2$ and,
being the Cosmological asymmetries generated through leptogenesis linear in
the CPV asymmetries, the final result gets enhanced by the same factor.
Therefore, for any given set of couplings, one can always find an appropriate
rescaling such that the correct amount of Cosmological lepton asymmetry is
generated.  In practice, the rescaling factors $\kappa$ cannot be arbitrarily
large: first, they should respect the condition that all the fundamental
Yukawa couplings remain in the perturbative regime; second, as will be
discussed in section \ref{sec:lfv-processes}, the size of the $h$ couplings
(and thus also of the rescaling parameter $\kappa$) is also constrained by
experimental limits on lepton flavor violating decays.

\section{Boltzmann Equations}
\label{sec:BE}
\begin{figure}[t]
\begin{center}
\includegraphics[width=13.0cm,height=2.3cm,angle=0]{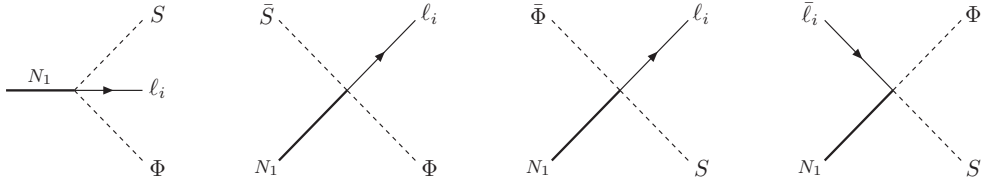}
\end{center}
\caption{Feynman diagrams for $1\leftrightarrow 3$ and
  $2\leftrightarrow 2$ $s$, $t$ and $u$ channel processes 
  after integrating out the heavy vectorlike fields $F_a$.}
 \label{fig:fig1}
 \end{figure}

In this section we compute the lepton asymmetry by solving the
appropriate BE. In general, to consistently derive
the evolution equation of the lepton asymmetry all the possible
processes at a given order in the couplings have to be included. 
In the present case $1\leftrightarrow 3$ decays and inverse decays, and
$2\leftrightarrow 2$ $s$, $t$ and $u$ channel scatterings all occur at the
same order in the couplings and must be included altogether in the BE. The
Feynman diagrams for these processes are shown in Figure~\ref{fig:fig1}.  In
addition, the CPV asymmetries of some higher order multiparticle reactions
involving the exchange of one off-shell $N_1$, also contribute to the source
term of the asymmetries at the same order in the couplings than the CPV
asymmetries of decays and $2\leftrightarrow 2$ scatterings.  More precisely,
for a proper derivation of the BE it is essential that the CPV asymmetries 
of the off-shell  $3\leftrightarrow 3$ and
$2\leftrightarrow 4$ scattering processes depicted in figures \ref{fig:3to3a},
% \ref{fig:3to3b} 
and \ref{fig:2to4} in appendix \ref{sec:appendixA}, are
also taken into account. 
In order to do this, we follow Ref.~\cite{Nardi:2007jp} and we split 
the BE for the evolution of the density asymmetry of the  flavor $\ell_i$ as:   
\begin{equation}
  \label{eq:densities}
  \dot{Y}_{\Delta L_i} = 
  (\dot{Y}_{\Delta L_i})_{1\leftrightarrow 3} 
  + (\dot{Y}_{\Delta L_i})_{2\leftrightarrow 2} 
  + (\dot{Y}_{\Delta L_i})_{3\leftrightarrow 3}^{\mbox{\tiny{sub}}}
  + (\dot{Y}_{\Delta L_i})_{2\leftrightarrow 4}^{\mbox{\tiny{sub}}}\,,
\end{equation}
where $Y_{\Delta L_i}=(n_{\ell_i}- n_{\bar\ell_i})/s$ with $n_{\ell_i}$
($n_{\bar\ell_i}$) the number density of $\ell_i$ (anti)leptons, and $s$ the
entropy density.  The time derivative is defined as $\dot{Y}\equiv
s\,H\,z\,dY/dz$ where $z=M_{N_1}/T$ and $H$ is the Hubble parameter.
The first term on the r.h.s. of eq.~(\ref{eq:densities}) represents the
contribution of three body decays and inverse decays, the second term that of
$2\leftrightarrow 2$ scatterings, the third term is defined in terms 
of the off-shell (pole subtracted) $3\leftrightarrow 3$ multiparticle 
density rates $\gamma_{3\leftrightarrow 3}^{\mbox{\tiny{sub}}}= 
\gamma_{3\leftrightarrow 3}-\gamma_{3\leftrightarrow
  3}^{\mbox{\tiny{on-shell}}}$, and similarly for the fourth term.  
The BE for $Y_{\Delta_{\ell_i}}$ is derived by 
taking into account in  full the first two terms on the r.h.s., 
while for the remaining two terms only the corresponding CP asymmetry 
is important, since non-resonant contributions to the washouts from 
multiparticle processes are always negligible. 

As regards the equation for the evolution of the heavy neutrino density
$Y_{N_1}$,  only the diagrams in fig.~\ref{fig:fig1}, that are of leading order 
in the couplings,  are important.
We refer to appendix~\ref{sec:appendixA} for
a detailed derivation of the equations. The final result reads
\begin{align}
  \label{eq:BEforN-section}
  \dot{Y}_{N_1} &= 
  -\left(y_{N_1} - 1 \right) \gamma_{\text{tot}} \\
  \label{eq:BEforLA-section}
  \dot{Y}_{\Delta L_{i}} &=
  \left(y_{N_1} - 1 \right)\epsilon_{i}\gamma_{\text{tot}} - \Delta y_{i}
  \left[
    \gamma_{i} + \left(y_{N_1} - 1 \right)\gamma^{N_{1}\bar\ell_{i}}_{S\Phi}
  \right]\,, 
\end{align}
where in the last term of the second equation we have used the 
compact notation for the reaction densities 
$\gamma^{N_{1}\bar\ell_{i}}_{S\Phi} =\gamma(N_{1}\bar\ell_{i}\to {S\Phi})$.
 $\gamma_{i}$ and  $\gamma_{\text{tot}}$ are defined as 
\begin{eqnarray}
  \label{eq:rates}
    \gamma_{i} &=& 
 \gamma^{N_{1}}_{S\ell_{i}\Phi}+
  \gamma^{N_{1}\bar{S}}_{\Phi\ell_{i}} +
  \gamma^{N_{1}\bar{\Phi}}_{S\ell_{i}} + \gamma_{S\Phi}^{N_{1}\bar\ell_{i}} \\
\gamma_{\text{tot}} &=& \sum_{i=e,\mu,\tau} 
    \gamma_{i}+\bar\gamma_i, 
\end{eqnarray}
where in the second equation $\bar\gamma_i$ represents the 
sum of the  CP conjugates of the processes summed in $\gamma_{i}$.

 Since in this model $N_1$ decays are of the same order in the couplings
 than scatterings (that is ${\cal O}(\tilde\lambda^2)$), the appropriate condition
 that defines the {\it strong washout} regime in the case at hand reads:
\begin{equation}
  \label{eq:strong-washout}
  \left .\frac{\gamma_{\text{tot}}}{z\,H\,s}\right|_{z\sim 1}>1 \,\qquad {\rm
  (strong\ washout)}, 
\end{equation}
and conversely $\gamma_{\text{tot}}/(z\,H\,s)|_{z\sim 1}<1$ defines the {\it
  weak washout} regime.  Note that this is different from standard
leptogenesis, where at $z\sim 1$ two body decays generally dominate over
scatterings, and  e.g. the condition for the strong washout regime can be 
approximated as $\gamma_{\text{tot}}/(z\,H\,s)|_{z\sim 1}\sim
\Gamma_{N_1}/H|_{z\sim 1}>1$.

\section{Results}
\label{sec:results}
In this section we discuss a typical example of successful leptogenesis at the
scale of a few TeV.  The example presented is a general one. No particular
choice of the parameters has been performed, except for the fact that the low
energy neutrino data are reproduced within errors, and that the choice yields
an interesting washout dynamics well suited to illustrate how PFL works.  The
numerical value of the final lepton asymmetry ($Y_{\Delta L} \sim -5.4\times
10^{-10}$) is about a factor of 3 {\it larger} than what is indicated by
measurements of the Cosmic baryon asymmetry. This is however irrelevant since,
as was discussed in section 2., it would be sufficient a minor rescaling of
the couplings (or a slight change in the CPV phases) to obtain the precise
experimental result.  In the numerical analysis we have neglected the dynamics
of the heavier singlet neutrinos since the $N_\alpha$ masses are sufficiently
hierarchical to ensure that $N_{2,3}$ related washouts do not interfere with
$N_1$ dynamics. Moreover, in the (strong washout) fully flavored regime (that
is effective as long as $T < 10^{9}\,$GeV) the $N_{2,3}$ CPV asymmetries do not
contribute to the final lepton number asymmetry~\cite{Engelhard:2006yg}.

\begin{figure}[t!]
\begin{center}
% \subfloat[Fig. 4(a): The different $\gamma$'s involving $\ell_{\tau}$]
 {\includegraphics[width=6.5cm,height=5.5cm,angle=0]{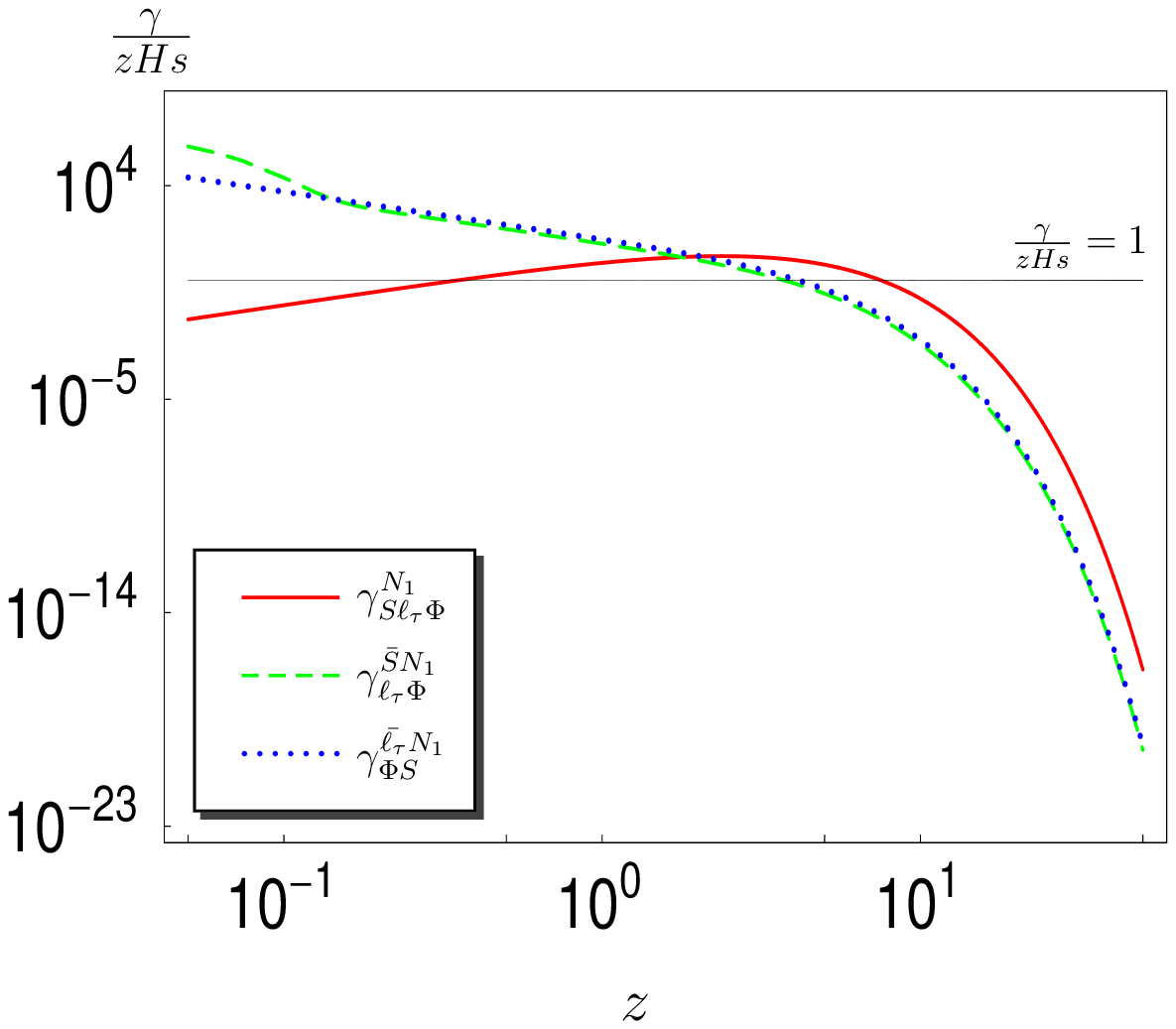}}
% \subfloat[Fig. 4(b): The different $\gamma$'s involving $\ell_{e}$]
\hspace{5mm}
 {\includegraphics[width=6.5cm,height=5.5cm,angle=0]{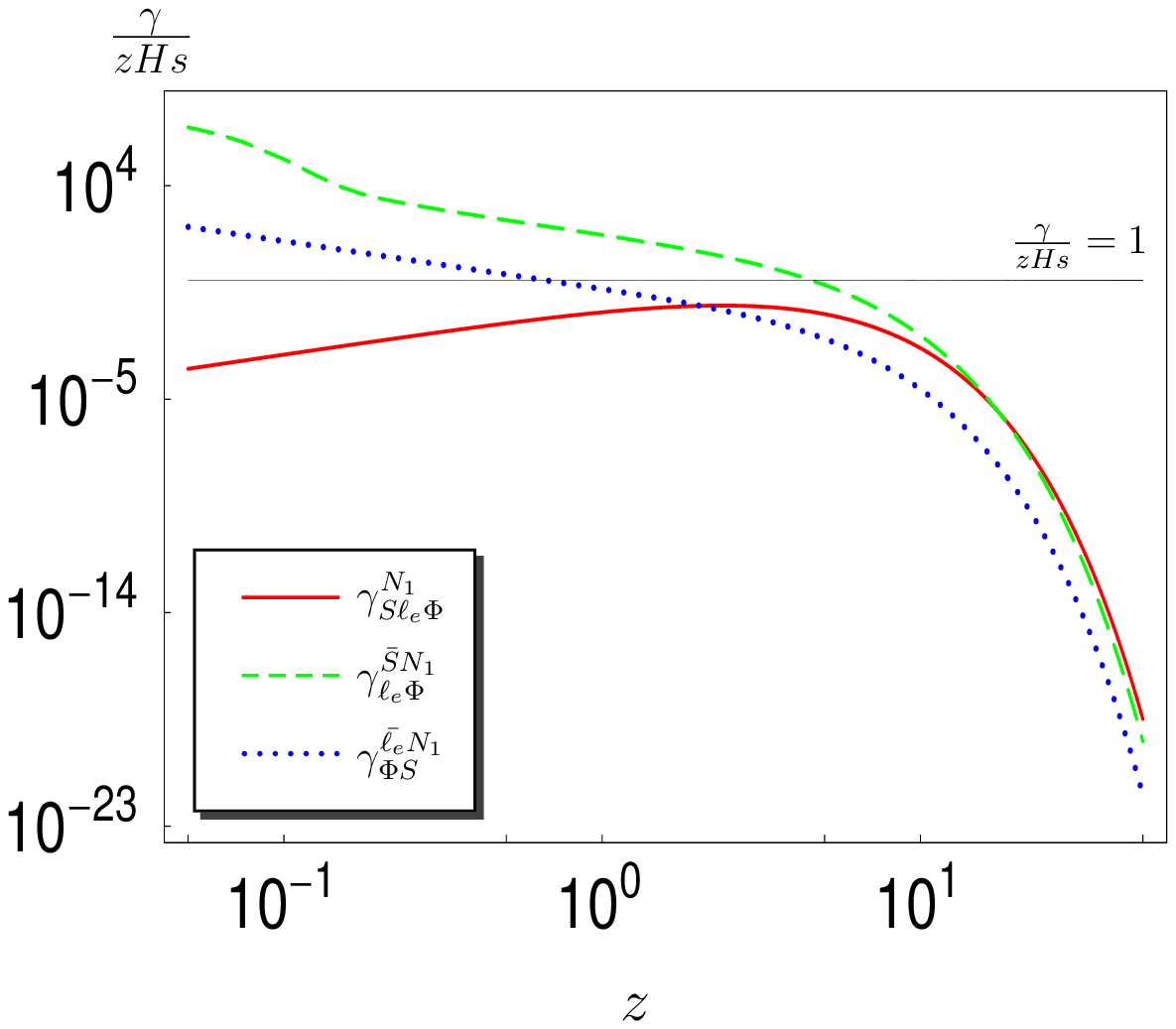}}
\end{center}
\caption{
Reaction densities  normalized to $zHs$  for  $N_1\to S\ell\Phi$ decays
(red solid lines),  $s$-channel $\bar S N_1 
\leftrightarrow \ell \Phi$ scatterings (green dashed lines), and 
$t,u$-channel scatterings in the point-like approximation (blue dotted lines).
Left panel:  $\tau$ flavor. Right panel:  electron flavor. 
}
\label{fig:fig4}
\end{figure}

\begin{figure}[t!!]
\begin{center}
%\subfloat[ Fig. 5(a): The total washout rates corresponding to the evolution]
{\includegraphics[width=6.5cm,height=5.5cm,angle=0]{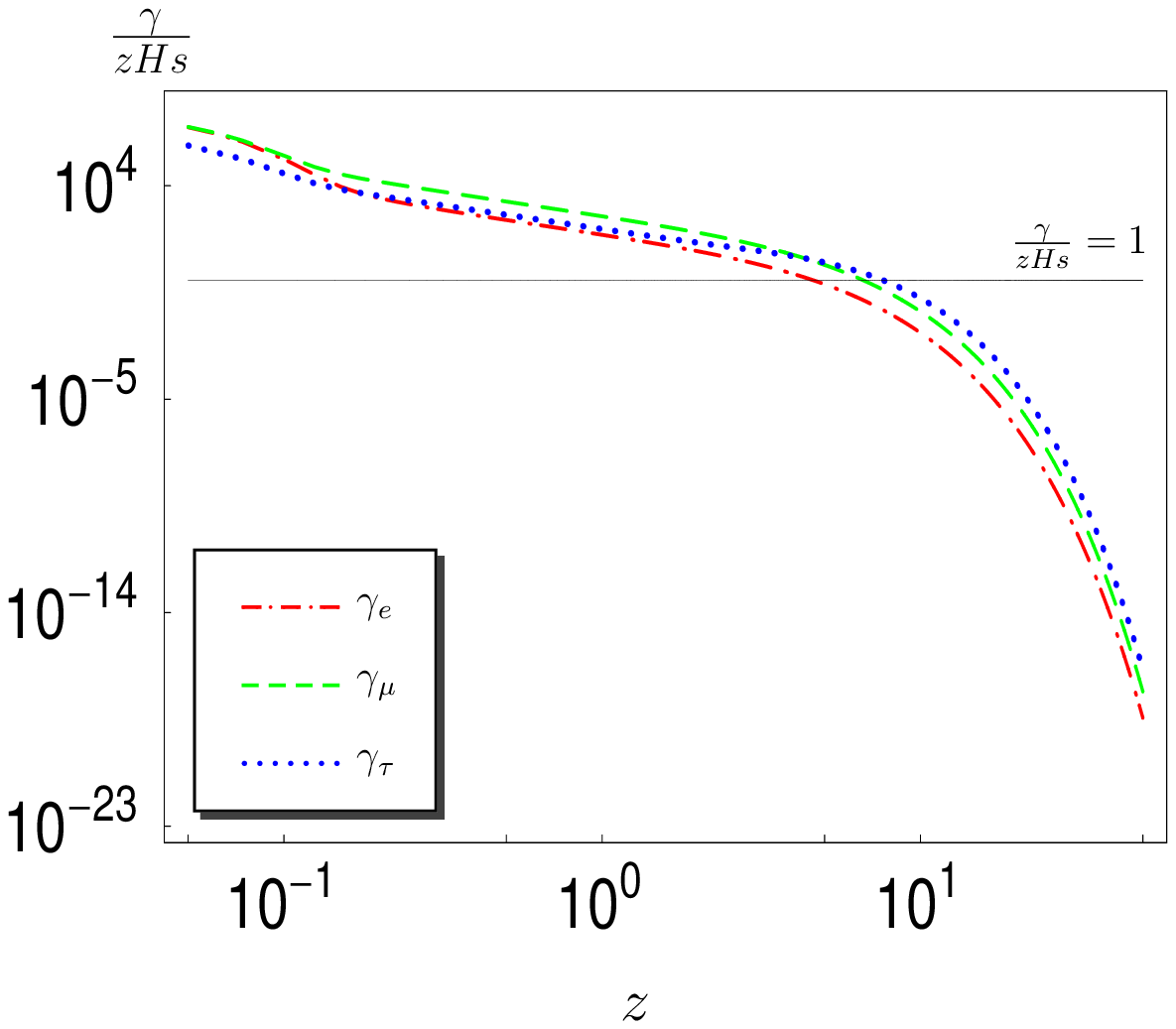}}
\hspace{5mm}
% \subfloat[ Evolution of the flavor asymmetries corresponding to the detailed]
{\includegraphics[width=6.5cm,height=5.5cm,angle=0]{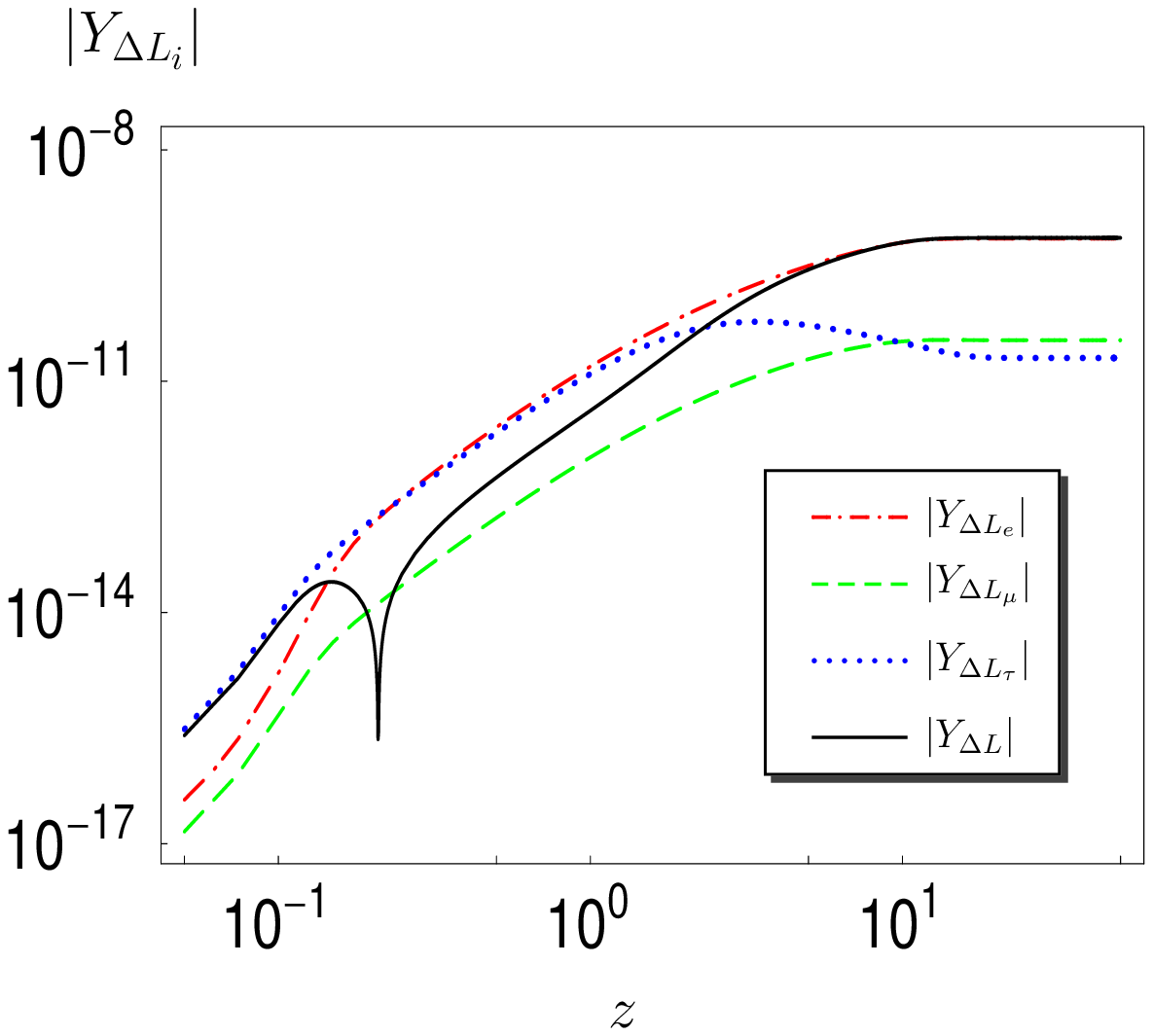}}
\end{center}
\caption{Left panel: the total washout rates for each lepton flavor 
  normalized to $zHs$ as a function of $z$.  Right panel: the
  evolution of the absolute value of the flavored density asymmetries and of
  the lepton number asymmetry (black solid line).  The flavor CPV asymmetries
  are $\epsilon_{1e} = -4.7\times 10^{-4}$, $\epsilon_{1\mu} = -1.9\times
  10^{-4}$ and $\epsilon_{1\tau} = 6.6\times 10^{-4}$.  The final values of
  the asymmetry densities (at $z\gg 1$) are $Y_{\Delta L_{e}} =-7.1\times
  10^{-10}$, $Y_{\Delta L_{\mu}} =-0.3\times 10^{-10}$,
  $Y_{\Delta L_{\tau}} =2.0\times 10^{-10}$. 
% $Y_{\Delta L}=  -5.4\times 10^{-10}$.  
}
\label{fig:fig5}
\end{figure}

In figure~\ref{fig:fig4} we show the behavior of the various reaction
densities for decays and scatterings, normalized to $sHz$, as a function of
$z=M_{N_1}/T$.  The results correspond to a mass of the lightest singlet
neutrino fixed to $M_{N_1}=2.5\,\text{TeV}$, the heavier neutrino masses are $
M_{N_2}=10\,$TeV and $M_{N_3}=15\,$TeV, and the relevant mass ratios
$r_a=M_{N_1}/M_{F_a}$ for the messenger fields are $r_{1,2,3}=0.1,0.01,0.001$
(the effects of the lightest $F$ resonances can be seen in the $s$-channel
rates in both panels in fig.~\ref{fig:fig4}).  The fundamental Yukawa
couplings $h$ and $\lambda$ are chosen to satisfy the requirement that the
seesaw formula eq.~(\ref{eq:nmm}) reproduces within $2\,\sigma$ the low energy
data on the neutrino mass squared differences and mixing
angles~\cite{Schwetz:2008er}.  Typically, when this requirement is fulfilled,
one also ends up with a dynamics for all the lepton flavors in the strong
washout regime. This is shown in the left panel in fig.~\ref{fig:fig5}
where we present the total rates for the three flavors.

The left panel in fig.~\ref{fig:fig4} refers to the decay and scattering rates
involving the $\tau$-flavor that, in our example, is the flavor more strongly
coupled to $N_1$, and that thus suffers the strongest washout. It is worth
noticing that, due to the fact that in this model scatterings are not
suppressed by additional coupling constants with respect to the decays, the
decay rate starts dominating the washouts only at $z\gsim 1$.  The right panel
in fig.~\ref{fig:fig4} depicts the reaction rates for the electron flavor,
that is the more weakly coupled, and for which the strong washout condition
eq.~(\ref{eq:strong-washout}) is essentially ensured by sizeable $s$-channel
scatterings.  Scatterings and decay rates for the $\mu$-flavor are not shown,
but they are in between the ones of the previous two flavors.

The total reaction densities that determine the washout rates for the
different flavors are shown in the first panel in figure~\ref{fig:fig5}.  The
evolution of these rates with $z$ should be confronted with the evolution of
(the absolute value of) the asymmetry densities for each flavor, depicted in
the second panel on the right. Since, as already stressed several times, PFL
is defined by the condition that the sum of the flavor CPV asymmetry vanishes
($\sum_j \epsilon_{1j}=0$), it is the hierarchy between these washout rates
that in the end is the responsible for generating a net lepton number
asymmetry.  In the case at hand, the absolute values of the flavor CPV
asymmetries satisfy the condition $|\epsilon_\mu| < |\epsilon_e| <
|\epsilon_\tau|$, as can be inferred directly by the fact that at $z< 0.1$,
when the effects of the washouts are still negligible, the asymmetry densities
satisfy this hierarchy.  Moreover, since $\epsilon_{\mu,e} < 0$ while
$\epsilon_\tau>0$, initially the total lepton number asymmetry, that is
dominated by $Y_{\Delta L_{\tau}}$, is positive.  As washout effects become
important, the $\tau$-related reactions (blue dotted line in the left panel)
start erasing $Y_{\Delta L_{\tau}}$ more efficiently than what happens for
the other two flavors, and thus the initial positive asymmetry is driven
towards zero, and eventually changes sign around $z=0.2$. This change of sign
corresponds to the steep valley in the absolute value $|Y_{\Delta L}|$ that is
drawn in the figure with a black solid line.  
Note that when all flavors are in  the strong washout regime, 
as in the present case, the condition for the occurrence of 
this `sign inversions'  is simply given by  
$ {\rm max_{j\in e,\mu}}
\left(|\epsilon_j|/|\tilde\lambda_{1j}|^2\right)
\gsim \epsilon_\tau/|\tilde\lambda_{1\tau}|^2$. 
From this point onwards, the
asymmetry remains negative, and since the electron flavor is the one that
suffers the weakest washout, $Y_{\Delta L_{e}}$ ends up dominating all the
other density asymmetries.  In fact, as can be seen from the 
right panel in fig.~\ref{fig:fig5}, it is  $Y_{\Delta L_{e}}$
that determines to a large extent the final value of the lepton asymmetry 
$Y_{\Delta  L}=-5.4\times 10^{-10}$.

A few comments are in order regarding the role played by the $F_a$ fields.
Even if $M_{N_1}\ll M_{F_a}$, at large temperatures $z\gg 1$ the tail of the
thermal distributions of the $N_1,\,S$ and $\Phi$ particles allows the
on-shell production of the lightest $F$ states. A possible asymmetry generated
in the decays of the $F$ fields can be ignored for two reasons: first because
due to the rather large $h$ and $\lambda$ couplings $F$ decays occur to a good
approximation in thermal equilibrium, ensuring that no sizeable asymmetry can
be generated, and second because the strong washout dynamics that
characterizes $N_1$ leptogenesis at lower temperatures is in any case
insensitive to changes in the initial conditions.

In conclusion, it is clear from the results of this section that the model
encounters no difficulties to allow for the possibility of generating the
Cosmic baryon asymmetry at a scale of a few TeVs. Moreover, our analysis
provides a concrete example of PFL, and shows that the condition
$\epsilon_1\neq 0$ is by no means required for successful leptogenesis.

\section{Lepton Flavor Violating Decays}
\label{sec:lfv-processes}

We have seen that a particular feature of this model is that the rescaling of
the couplings eq.~(\ref{eq:coupling-rescaling-gen}) can enhance the CPV
asymmetry by a factor $\kappa^2$ (see eq.~(\ref{eq:rescaled-CPV-asymm}))
without affecting neither the low energy neutrino physics nor the washout
rates.  In practice, it is this decoupling of the CPV asymmetries from the
washouts that renders possible lowering the leptogenesis scale down to the
TeV.  It is then natural to ask how large the rescaling factor $\kappa$ can
be, or in other words how large the $h$ couplings can become, without
incurring in the violation of some phenomenological bound.  To this aim, in
this section we will derive upper bounds on the Yukawa couplings $h_{i a}$
from the non-observation of Lepton Flavor Violating (LFV) decays.  The set of
Yukawa interactions involving the heavy vectorlike fields $F_a$, the $SU(2)$
Higgs scalar $\Phi$ and lepton doublets $\ell_i$, can induce lepton flavor
violating radiative decays $\ell_i\to \ell_j\gamma$.  Here we will concentrate
on $\mu\to e\gamma$ that is the most constrained process.  Note that, with
respect to the radiative decays, lepton flavor violating decays like $\ell'\to
3 \ell$ are more suppressed since they are induced by box diagrams rather then
by penguin-type  diagrams. Therefore we will not consider them.

\begin{figure}[t]
  \centering
  \includegraphics[height=7cm,width=8cm]{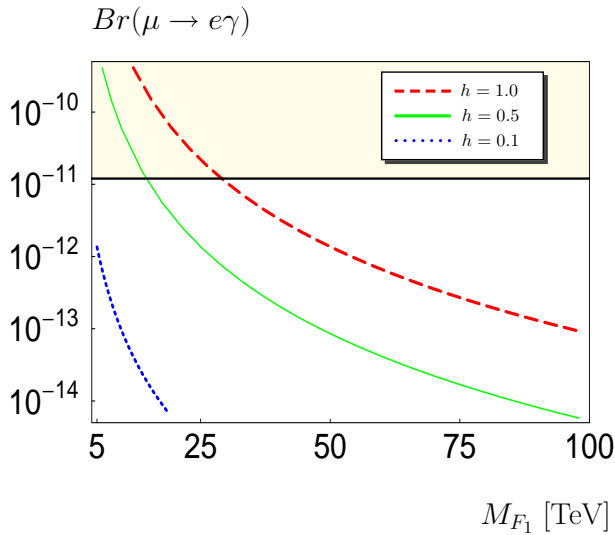}
  \caption{$Br(\mu\to e\gamma)$ as a function of the mass of the lightest
    vectorlike field $F_1$. No hierarchy for the $h$ couplings and a mild
    hierarchy for the $F$ masses ($M_{F_3}=1.3\cdot M_{F_2}=2\cdot M_{F_1}$)
    has been assumed. The region above the horizontal line is excluded by 
    current experimental limits. The hatched region corresponds to 
    non-perturbative couplings $h> \sqrt{4\pi}$. }
  \label{fig:mutoegamma}
\end{figure}

The partial decay width for the lepton flavor violating decay 
$\ell_i\to \ell_j\gamma$ reads 
\begin{equation}
  \label{eq:ljtoligamma}
  \Gamma(\ell_i\to \ell_j\gamma) = \frac{\alpha}{1024\pi^4}\frac{m_i^5}{M_W^4}
  \left|
    \sum_{a=1}^3 h_{i a}^* h_{j a} F(M_W^2/M_{F_a}^2)
  \right|^2\,.
\end{equation}
Here $M_W$ is the $W^\pm$ $SU(2)$ gauge boson mass which enter the loop
through its longitudinal component $\Phi^\pm$
and $m_i$ is the mass of the decaying lepton (the mass of the 
final state lepton $m_j$ has been neglected). 
$F(x)$ is a loop function given by
\begin{equation}
  \label{eq:loop-function}
  F(x) = \frac{x}{12(1-x)^4}
  \left(
    2 + 3x - 6x^2 + x^3 + 6x\log x  
  \right)
\end{equation}
and for $x\to 0$, $F(x)\to x/6$. Since $M_{F_a}\gg M_W$, $F(M_W^2/M_{F_a}^2)$
strongly suppresses the LFV radiative decays and thus, in general, the Yukawa
couplings $h_{i a}$ will not be strongly constrained by the current
experimental upper limits.  In fact, as is shown in fig.~\ref{fig:mutoegamma},
for $M_F>$ a few TeVs and Yukawa couplings $\lsim .5$, the radiative LFV decay
rates always remain far below the present limits.  Only for $h\gsim 2$ (that
is close to the perturbative limit represented by the hatched region in
fig.~\ref{fig:mutoegamma}) and $M_{F_1}\sim 25\,$TeV (that was our choice in
the numerical example) the $\mu\to e\gamma$ decay rate becomes comparable to
the present sensitivity, and is well within the reach of future
experiments~\cite{meg}. Thus we can safely conclude that 
present limits on LFV decays do not place any serious constraint 
on the viability of TeV scale leptogenesis within the PFL model 
discussed in this paper.

\section{Conclusions}
\label{sec:conclusions}
Variations of the standard leptogenesis scenario can arise from the
presence of additional (flavor) symmetries broken around or below the
scale at which lepton number is effectively broken.  Quite
generically, the resulting scenarios can yield new qualitative and
quantitative changes on the way leptogenesis is realized. Here we have
considered a model in which an Abelian $U(1)_X$ flavor symmetry, still
unbroken during the leptogenesis era, is added to the SM gauge
symmetry group.  We have also assumed that the messengers fields
responsible for the effective mass operators of the light particles
are heavier that the lightest Majorana neutrino $N_1$ and thus, during
leptogenesis, cannot be produced.  The model has the remarkable
feature that the total CPV asymmetry in $N_1$ decays vanishes, while
the lepton flavor asymmetries are generically nonvanishing, and thus
it constitutes an explicit realization of the scenario that we have
called {\it purely flavored leptogenesis}.

By using the BE specific for this model, we have studied the evolution
of the asymmetry densities for the different lepton flavors as the
temperature changes, and we have found that successful leptogenesis
can occur at a scale as low as a few TeVs. This possibility is due to
the fact that the size of the flavored CPV asymmetries is decoupled
from the strength of the washouts and from low energy neutrino
physics.  This allows to rescale the CPV asymmetries up to rather
large values, leaving unaffected the washout rates as well as the
light neutrino masses and mixings.  Our model shows that if new
unbroken symmetries are present at a scale below the leptogenesis
scale, this could have a very interesting and even surprising impact
on the way leptogenesis is realized.

 \section{Acknowledgments}
 \label{sec:acknowledgments}
 LAM would like to thank short term training fellowships from
 the EU HELEN program and hospitality from the INFN-LNF Theory Group 
 in Frascati (Rome). The work of EN is supported in part by
 Colciencias under contract 1115-333-18739.

 % ------------
 \appendix
 % ------------
 \section{Boltzmann Equations}
 \label{sec:appendixA}
 Following mainly ref.~\cite{Nardi:2007jp}, we start by introducing a set of
 compact notations.  We normalize particle densities to the equilibrium
 densities, $y_a\equiv Y_a/Y_a^{\text{eq}}$ where $Y_a=n_a/s$ with $n_a$ the
 particle number density and $s$ the entropy density, and we define the time
 derivative as $\dot Y = sHz\,dY/dz$. Reaction densities are denoted by
 $\gamma_B^A$ where $A$ and $B$ are respectively the initial and final states
 of the specific decay or scattering process.  
% For the collision terms we adopt the shorthand notation
 %
% \begin{equation}
%   \label{eq:shorthand-collisionterm}
%[A\leftrightarrow B] = \left(\prod_i y_{a_i}\right) \gamma^A_B - 
% \left(\prod_j y_{b_j}\right) \gamma^B_A\,.
%  \end{equation}
 %
As in eq.~(\ref{eq:densities}), 
we divide  he  BE for the evolution of the density asymmetry 
 of the flavor $\ell_i$ 
${Y}_{\Delta L_{i}} \equiv Y_{\ell_i}-Y_{\bar\ell_i}$ 
into different contributions 

 \begin{equation}
   \label{eq:densitiesA}
   \dot{Y}_{\Delta L_{i}} = 
   (\dot{Y}_{\Delta L_{i}})_{1\leftrightarrow 3} 
   + (\dot{Y}_{\Delta L_{i}})_{2\leftrightarrow 2} 
   + (\dot{Y}_{\Delta L_{i}})_{3\leftrightarrow 3}^{\mbox{\tiny{sub}}}
   + (\dot{Y}_{\Delta L_{i}})_{2\leftrightarrow 4}^{\mbox{\tiny{sub}}}\,, 
 \end{equation}
and we derive the explicit form of the different contributions in
the following sections.

 \subsection{$1\leftrightarrow 3$ and $2\leftrightarrow 2$ processes}

  The contributions $(\dot{Y}_{\Delta L_{i}})_{1\leftrightarrow 3}$
  and $(\dot{Y}_{\Delta L_{i}})_{2\leftrightarrow 2}$ 
 in eq.~(\ref{eq:densitiesA}) arise from the 
  $1\leftrightarrow 3$ and $2\leftrightarrow 2$ reactions  depicted in
  figure \ref{fig:fig1}, that are all of ${\cal O}(\lambda^2 h^2)$. 
   Using CPT invariance
  ($\gamma^A_B=\gamma^{\bar B}_{\bar A}$) and after linearizing in the
  CPV asymmetries and in the asymmetry densities $\Delta y_{a}\equiv  y_{a}-
  y_{\bar a}$ 
  they can be written as:
 \begin{align}
   \label{Arel1} 
   (\dot{Y}_{\Delta L_{i}})_{N_1\leftrightarrow \Phi S\ell_i} &=
   (y_{N_{1}} + 1)\Delta\gamma^{N_{1}}_{S\ell_{i}\Phi} - 
   (\Delta y_{S} + \Delta y_{\Phi} + \Delta y_{\ell_{i}}) 
   \gamma^{N_{1}}_{S\ell_{i}\Phi},
   \\
   \label{Arel2}
   (\dot{Y}_{\Delta L_{i}})_{N\bar S\leftrightarrow \Phi \ell_i} &= 
   (y_{N_{1}} + 1)\Delta\gamma^{N_{1}\bar{S}}_{\Phi\ell_{i}} -
   (y_{N_{1}}\Delta y_{S} + \Delta y_{\Phi} + \Delta y_{\ell_{i}})
   \gamma^{N_{1}\bar{S}}_{\Phi\ell_{i}},
   \\
   \label{Arel3}
   (\dot{Y}_{\Delta L_{i}})_{N\bar\Phi\leftrightarrow S\ell_i} &= 
   (y_{N_{1}} + 1)\Delta\gamma^{N_{1}\bar{\Phi}}_{S\ell_{i}} -
   (\Delta y_{S}+y_{N_{1}}\Delta y_{\Phi} + \Delta y_{\ell_{i}})
   \gamma^{N_{1}\bar{\Phi}}_{S\ell_{i}},
   \\
   \label{Arel4}
   (\dot{Y}_{\Delta L_{i}})_{N\bar\ell_i\leftrightarrow \Phi S} &= 
   (y_{N_{1}}+1)\Delta\gamma_{S\Phi}^{N_{1}\bar{\ell_{i}}} - 
   (\Delta y_{S} + \Delta y_{\Phi} + y_{N_{1}}
   \Delta y_{\ell_{i}})\gamma_{S\Phi}^{N_{1}\bar{\ell_{i}}}\,,
 \end{align}
 where $(\dot{Y}_{\Delta \ell_{i}})_{N\bar S\leftrightarrow \Phi \ell_i}$,
 $(\dot{Y}_{\Delta L_{i}})_{N\bar\Phi\leftrightarrow S\ell_i}$ and
 $(\dot{Y}_{\Delta L_{i}})_{N\bar\ell_i\leftrightarrow \Phi S}$ are the $s$,
 $t$ and $u$ channel contributions to the $2\leftrightarrow 2$ scattering term
 $(\dot{Y}_{\Delta L_{i}})_{2\leftrightarrow 2}$.  For completeness, in these
 equations as well as in the following we keep trace of the density
 asymmetries $\Delta y_{\Phi}$ and $\Delta y_{S}$ of the Higgs and of the
 $S$-scalar. This is needed if one wishes to take into account spectator
 processes~\cite{Nardi:2005hs}.  The related effects (that can be as large as
 40\% ~\cite{Nardi:2005hs}) depend, however, on the specific interactions of
 $S$ with other particles (quarks) and are thus model dependent, and have been
 neglected in the present analysis.  After summing up
 eqs.~(\ref{Arel1}-\ref{Arel4}) we obtain
 \begin{align}
   \label{eq:1to3and2to2-complete}
   (\dot{Y}_{\Delta L_{i}})_{1\leftrightarrow 3}  + 
  (\dot{Y}_{\Delta L_{i}})_{2\leftrightarrow 2}  
 &
   = (y_{N_{1}}+1) (\Delta\gamma^{N_{1}}_{S\ell_{i}\Phi} + 
   \Delta\gamma^{N_{1}\bar{S}}_{\Phi\ell_{i}} +
   \Delta\gamma^{N_{1}\bar{\Phi}}_{S\ell_{i}} +
   \Delta\gamma_{S\Phi}^{N_{1}\bar\ell_{i}})
   \nonumber\\
   &
   - \Delta y_{\ell_{i}}( \gamma^{N_{1}}_{S\ell_{i}\Phi} +
   \gamma^{N_{1}\bar{s}}_{\Phi\ell_{i}}+\gamma^{N_{1}\bar{\Phi}}_{s\ell_{i}} +
   y_{N_{1}}\gamma_{S\Phi}^{N_{1}\bar\ell_{i}})\nonumber\\
   &
   - \Delta y_{\Phi}( \gamma^{N_{1}}_{S\ell_{i}\Phi} +
   \gamma^{N_{1}\bar{S}}_{\Phi\ell_{i}} +
   y_{N_{1}}\gamma^{N_{1}\bar{\Phi}}_{S\ell_{i}} +
   \gamma_{S\Phi}^{N_{1}\bar\ell_{i}})\nonumber\\
   &
   - \Delta y_{S}( \gamma^{N_{1}}_{S\ell_{i}\Phi}+y_{N_{1}}
   \gamma^{N_{1}\bar{S}}_{\Phi\ell_{i}} +
   \gamma^{N_{1}\bar{\Phi}}_{S\ell_{i}} + \gamma_{S\Phi}^{N_{1}\bar\ell_{i}})\,.
 \end{align}
As is usual when only lowest order processes are included in the
 BE, the factor $(y_{N_{1}}+1)$ signals an incorrect thermodynamical
 behavior (generation of an asymmetry in thermal equilibrium).  In
 order to get the correct result we need to include in the BE also the
 CPV asymmetries of higher order processes, like the $3\leftrightarrow 3$ and $2\leftrightarrow 4$
 scatterings in which one $N_1$ is exchanged as a virtual state in the
 internal lines.  This is carried out in the following two sections.
 %

 %
 % \begin{figure}[h!!]
 %  \begin{center}
 %    \includegraphics[width=9.5cm,height=2.5cm,angle=0]{3to3b.eps}
 %  \end{center}
 %  \caption{Feynman diagrams for the  $3\leftrightarrow 3$ processes: 
 %    $S\ell_{i}S\leftrightarrow\bar\Phi\bar \ell_{i}\bar\Phi$,
 %    $\Phi S S\leftrightarrow \bar\ell_{i}\bar\ell_{i}\Phi$  
 %    and $\Phi S \Phi \leftrightarrow \bar\ell_{i}\bar\ell_{i}\bar S$.}
 %  \label{fig:3to3b}
 % \end{figure}

 \subsection{$3\leftrightarrow 3$ processes}
\begin{figure}[t]
   \begin{center}
     \includegraphics[width=9.50cm,height=5.0cm,angle=0]{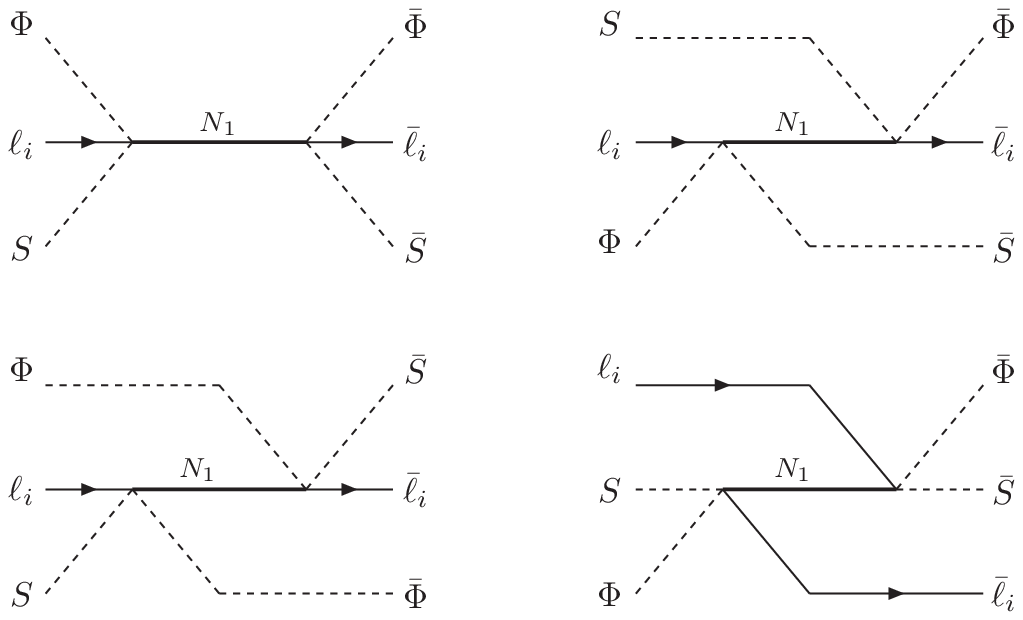}\\
 \vspace{.5cm}
    \includegraphics[width=9.5cm,height=2.5cm,angle=0]{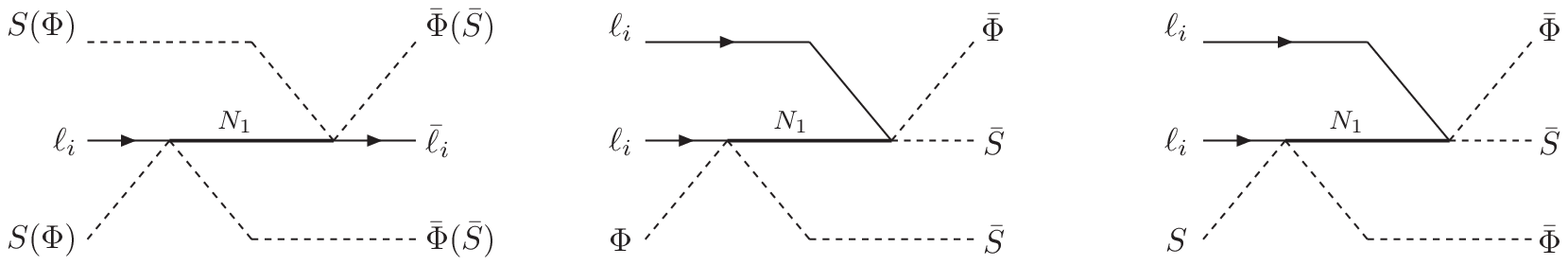}
   \end{center}
   \caption{Feynman diagrams for   $3\leftrightarrow 3$  scattering
     processes yielding $|\Delta L_{i}|=2$.  $\Phi\ell_{i}
     S\leftrightarrow\bar\Phi\bar\ell_{i}\bar S$ (first four diagrams);
     $S\ell_{i}S\leftrightarrow\bar\Phi\bar \ell_{i}\bar\Phi$ (and $\Phi
     \ell_{i}\Phi \leftrightarrow\bar S\bar \ell_{i}\bar S$) (fifth diagram);
     $\ell_{i}\ell_{i}\Phi \leftrightarrow \bar\Phi \bar S \bar S $ and
     $\ell_{i}\ell_{i} S \leftrightarrow \bar\Phi \bar \Phi \bar S $ (last two
     diagrams).  }
   \label{fig:3to3a}
 \end{figure}

 Multiparticle tree level processes in which one $N_1$ is exchanged in one
 internal line can be divided into on-shell and off-shell parts.  For the
 on-shell parts, when the $N_1$ line in the amplitude is cut, we obtain either
 the $1\leftrightarrow 3$ or the $2\leftrightarrow 2$ diagrams in
 fig.~\ref{fig:fig1} which were already accounted for in
 eq.~(\ref{eq:1to3and2to2-complete}).  We then have to consider only the
 off-shell contributions, denoted as $\gamma'$, where the superscript is a
 reminder that the given reaction has the on-shell piece subtracted out.  The
 contribution of these off-shell processes to the washouts is negligible;
 however, the contribution of their CPV asymmetries cannot be neglected.
%
% \begin{equation}
%   \label{eq:3to3contributions}
%   (\dot Y_{\Delta \ell_i})_{3\leftrightarrow 3}^{\mbox{\tiny{sub}}} = [3\leftrightarrow 3]'\,
% \end{equation}
 %
% where $[3\leftrightarrow 3]'$ refers to the $3\leftrightarrow 3$ 
% reactions with the on-shell piece subtracted out. 
 Figure~\ref{fig:3to3a} shows the set of Feynman diagrams for
 $3\leftrightarrow 3$ scattering processes yielding $|\Delta L_{i}|=2$. They
 are of two types: $\ell_i \leftrightarrow \bar \ell_i$ and $ \ell_i\ell_i
 \leftrightarrow 0 $.  We do not show the analogous $|\Delta L_{i}|=1$
 diagrams, since they can have either one $\bar \ell_j$ or one $\ell_j$
 ($j\neq i$) attached to one external leg, and for this reason their number is
 rather large.
 Since the CPV asymmetries of the full  processes are of higher
 order in the couplings, at ${\cal O}(\lambda^2 h^4)$ the asymmetries of the
 off-shell parts are equal in magnitude and opposite in sign to the CPV 
 asymmetries of their on-shell parts. In turn, the latter's are directly
 related to the CPV asymmetry in $N_1$ decays and scatterings.  Denoting the on
 shell parts of the rates as $\gamma^{(\text{on})}$, we then have:
 \begin{equation}
   \label{eq:off-shell-on-shel}
   \Delta \gamma'^{A}_B = -\Delta \gamma^{(\text{on})\,A}_B\,.
 \end{equation}
 The on-shell pieces of the $3\leftrightarrow 3$ reactions for $|\Delta L_{i}|=2$
processes can be written as
 \begin{align}
   \label{eq:onshell-reactionD}
   \gamma^{(\text{on})\, \Phi\ell_{i} S}_{\bar\Phi\bar\ell_{i}\bar S} &=
    \gamma^{\Phi\ell_{i} S}_{N_1}  P^{N_1}_{\bar\Phi\bar\ell_i\bar S} 
   +\gamma^{\Phi\ell_i}_{\bar S N_1} P^{S N_1}_{\bar\Phi\bar\ell_i} 
   + \gamma^{S\ell_i}_{\bar\Phi N_1}P^{\Phi N_1}_{\bar S\bar\ell_i} 
   + \gamma^{S\Phi}_{\bar\ell_i N_1} P^{\ell_i N_1}_{\bar S\bar\Phi},\\ 
   \gamma^{(\text{on})\, S\ell_i S}_{\bar\Phi\bar\ell_i\bar\Phi} &=
   \gamma^{S\ell_i}_{\bar\Phi N_1} P^{S N_1}_{\bar\ell_i\bar\Phi}, \\
   \gamma^{(\text{on})\, \Phi \ell_i \Phi }_{\bar S\bar\ell_i\bar S} &=
   \gamma^{\Phi \ell_i}_{\bar S N_1} P^{\Phi  N_1}_{\bar\ell_i\bar S}, \\
   \gamma^{(\text{on})\, \Phi S S}_{\bar\ell_i\bar\ell_i\bar\Phi} &=
   \gamma^{\Phi S}_{\bar\ell_i N_1} P^{S N_1}_{\bar\ell_i\bar\Phi},\\
   \gamma^{(\text{on})\, \Phi S \Phi}_{\bar\ell_i\bar\ell_i\bar S} &=
  \gamma^{\Phi S}_{\bar\ell_i N_1} P^{\Phi N_1}_{\bar\ell_i\bar S} \,.
   \label{eq:onshell-reactionS}
 \end{align}
 The $P^A_B$ represents the (decay or scattering) probability for the
 transition $A\to B$, and were introduced in \cite{Nardi:2007jp}
 to generalize the zero temperature branching ratios to the finite temperature
 case, when the $N_1$'s have a finite probability to scatter inelastically
 with particles in the plasma before decaying.  We have for example:
 \begin{equation}
   \label{eq:probability}
   P^{N_1}_{\bar\Phi\bar\ell_i\bar S} =
 \frac{\gamma^{N_1}_{\bar\Phi\bar\ell_i\bar S}}{\gamma_{\text{tot}}}, \qquad 
   P^{S N_1}_{\bar\Phi\bar\ell_i} =
 \frac{\gamma^{S N_1}_{\bar\Phi\bar\ell_i}}{\gamma_{\text{tot}}}, 
 \qquad {\rm etc.}, 
 \end{equation}
 where
\begin{equation}
\label{eq:gammatot}
\gamma_{\text{tot}}=\sum_i\left( \gamma_{S\Phi\ell_i}^{N_1} +
   \gamma_{\bar S \bar\Phi \bar\ell_i}^{N_1} + \gamma_{\ell_i \Phi}^{\bar S
     N_1} + \gamma_{\bar\ell_i \bar\Phi}^{S N_1} + \gamma_{\ell_i S}^{\bar\Phi
     N_1} + \gamma_{\bar\ell_i \bar S}^{\Phi N_1} + \gamma_{\Phi S}^{\ell_i
     N_1} + \gamma_{\bar\Phi \bar S}^{\bar\ell_i N_1}\right)\,. 
\end{equation}
Using the set of equations 
(\ref{eq:onshell-reactionD}-\ref{eq:onshell-reactionS}) together with the
corresponding CP conjugate equations, including the contributions from
$|\Delta L_{i}|=1$ processes, and defining ${\cal P}^a_b
\equiv P^a_b + P^{\bar a}_{\bar b}$, the different contributions to $(\dot
Y_{\Delta\ell_i})_{3\leftrightarrow 3}^{\mbox{\tiny{sub}}}$ can be written as:
 \begin{align}
\label{eq:3to3-complete}
   %\label{eq:3to3-ind-contributions}
   % --------- 1 ----------
(\dot{Y}_{\Delta L_{i}})_{\Phi\ell_{i}
  S\leftrightarrow\bar{\Phi}\bar{\ell_{j}}\bar{S}} &= -
\Delta\gamma^{N_{1}}_{S\Phi\ell_{i}}\sum_{j}{\cal P}^{N_{1}}_{S\Phi\ell_{j}} -
\Delta\gamma^{N_{1}\bar{\Phi}}_{S\ell_{i}}\sum_{j}{\cal
  P}^{N_{1}\bar{\Phi}}_{S\ell_{j}}
\nonumber\\
& \quad - \Delta\gamma^{N_{1}\bar{S}}_{\Phi\ell_{i}}\sum_{j}{\cal
  P}^{N_{1}\bar{S}}_{\Phi\ell_{j}}
- \Delta\gamma^{\bar{\Phi}\bar{S}}_{N_{1}
\ell_{i}}\sum_{j}{\cal P}^{N_{1}\bar{\ell}_{j}}_{\Phi S}\,,\nonumber\\
% --------- 2 ----------
(\dot{Y}_{\Delta \ell_{i}})_{\Phi\ell_{i} S\leftrightarrow\Phi\ell_{j}S}
&= - \Delta\gamma^{N_{1}}_{S\Phi\ell_{i}}
\sum_{j}{\cal P}^{N_{1}}_{S\Phi\ell_{j}}\,,\nonumber\\
% --------- 3 ----------
(\dot{Y}_{\Delta \ell_{i}})_{\Phi\ell_{i}
  \bar{S}\leftrightarrow\Phi\ell_{j}\bar{S}}
&=- \Delta\gamma^{N_{1}\bar{S}}_{\Phi\ell_{i}}
\sum_{j}{\cal P}^{N_{1}\bar{S}}_{\Phi\ell_{j}}\,,\nonumber\\
% --------- 4 ----------
(\dot{Y}_{\Delta \ell_{i}})_{\bar{\Phi}\ell_{i}
  S\leftrightarrow\bar{\Phi}\ell_{j}S}
&=- \Delta\gamma^{N_{1}\bar{\Phi}}_{S\ell_{i}}
\sum_{j}{\cal P}^{N_{1}\bar{\Phi}}_{S\ell_{j}}\,,\nonumber\\
% --------- 5 ----------
(\dot{Y}_{\Delta \ell_{i}})_{\bar{\Phi}\ell_{i}
  \bar{S}\leftrightarrow\bar{\Phi}\ell_{j}\bar{S}}
&=-\Delta\gamma^{\bar{\Phi}\bar{S}}_{N_{1}\ell_{i}}
\sum_{j}{\cal P}^{N_{1}\bar{\ell}_{j}}_{\Phi S}\,,\nonumber\\
% --------- 6 ----------
(\dot{Y}_{\Delta L_{i}})_{S\ell_{i}
  S\leftrightarrow\bar{\Phi}\bar{\ell}_{j}\bar{\Phi}}
&=- \Delta\gamma^{N_{1}\bar{\Phi}}_{S\ell_{i}}
\sum_{j}{\cal P}^{N_{1}\bar{S}}_{\Phi\ell_{j}} \,,\nonumber\\
% --------- 7 ----------
(\dot{Y}_{\Delta
  L_{i}})_{\Phi\ell_{i}\Phi\leftrightarrow\bar{S}\bar{\ell}_{j}\bar{S}}
&=- \Delta\gamma^{N_{1}\bar{S}}_{\Phi\ell_{i}}
\sum_{j}{\cal P}^{N_{1}\bar{\Phi}}_{S\ell_{j}} \,,\nonumber\\
% --------- 8 ----------
(\dot{Y}_{\Delta L_{i}})_{\ell_{i} \ell_{j}
  \Phi\leftrightarrow\bar{S}\bar{\Phi}\bar{S}}
&=-\Delta\gamma^{N_{1}\bar{S}}_{\Phi\ell_{i}}
\sum_{j}{\cal P}^{N_{1}\bar{\ell}_{j}}_{\Phi S}-
\Delta\gamma^{\bar{\Phi}\bar{S}}_{N_{1}\ell_{i}}
\sum_{j}{\cal P}^{N_{1}\bar{S}}_{\Phi\ell_{j}}\,,\nonumber\\
% --------- 9 ----------
(\dot{Y}_{\Delta L_{i}})_{\ell_{i} \ell_{j}
  S\leftrightarrow\bar{\Phi}\bar{S}\bar{\Phi}}
&=-\Delta\gamma^{N_{1}\bar{\Phi}}_{S\ell_{i}}
\sum_{j}{\cal P}^{N_{1}\bar{\ell}_{j}}_{\Phi S}-
\Delta\gamma^{\bar{\Phi}\bar{S}}_{N_{1}\ell_{i}}
\sum_{j}{\cal P}^{N_{1}\bar{\Phi}}_{S\ell_{j}}\,,\nonumber\\
% --------- 10 ----------
(\dot{Y}_{\Delta L_{i}})_{S \ell_{i}
  \bar{S}\leftrightarrow\Phi\ell_{j}\bar{\Phi}}
&=-\Delta\gamma^{N_{1}\bar{\Phi}}_{S\ell_{i}}
\sum_{j}{\cal P}^{N_{1}\bar{S}}_{\Phi\ell_{j}}-
\Delta\gamma^{N_{1}\bar{S}}_{\Phi\ell_{i}}
\sum_{j}{\cal P}^{N_{1}\bar{\Phi}}_{S\ell_{j}}\,,\nonumber\\
% --------- 11 ----------
(\dot{Y}_{\Delta L_{i}})_{\ell_{i} \bar{\ell_{j}} \Phi\leftrightarrow
  S\Phi\bar{S}}
&=-\Delta\gamma^{N_{1}\bar{S}}_{\Phi\ell_{i}}
\sum_{j}{\cal P}^{N_{1}\bar{\ell}_{j}}_{\Phi S}-
\Delta\gamma^{\bar{\Phi}\bar{S}}_{N_{1}\ell_{i}}
\sum_{j}{\cal P}^{N_{1}\bar{S}}_{\Phi\ell_{j}}\,,\nonumber\\
% --------- 12 ----------
(\dot{Y}_{\Delta L_{i}})_{\ell_{i} \bar{\ell_{j}} S\leftrightarrow\Phi
  S\bar{\Phi}} &=-\Delta\gamma^{N_{1}\bar{\Phi}}_{S\ell_{i}}\sum_{j}{\cal
  P}^{N_{1}\bar{\ell}_{j}}_{\Phi
  S}-\Delta\gamma^{\bar{\Phi}\bar{S}}_{N_{1}\ell_{i}}\sum_{j}{\cal
  P}^{N_{1}\bar{\Phi}}_{S\ell_{j}},
 \end{align}
 where in the l.h.s., of these equations, whenever the $\ell_i \leftrightarrow
 \ell_j$ transition is involved, it is left understood that $i\neq j$. 
 To write the above relations, we have first completed the sums over
 flavor. For example for the second equation in (\ref{eq:3to3-complete}):
\begin{align}
  \label{eq:3to3-explanation}
% --------- 1 ----------
  (\dot{Y}_{\Delta \ell_{i}})_{\Phi\ell_{i} S\leftrightarrow\Phi\ell_{j}S}
  &= - \Delta\gamma^{N_{1}}_{S\Phi\ell_{i}}
\sum_{j\neq i}{\cal P}^{N_{1}}_{S\Phi\ell_{j}}+
{\cal P}^{N_{1}}_{S\Phi\ell_{i}}
\sum_{j\neq i}\Delta \gamma^{N_{1}}_{S\Phi\ell_{j}}
\nonumber\\
% --------- 2 ----------
 &
= - \Delta\gamma^{N_{1}}_{S\Phi\ell_{i}}
\sum_{j}{\cal P}^{N_{1}}_{S\Phi\ell_{j}}+
{\cal P}^{N_{1}}_{S\Phi\ell_{i}}
\sum_{j}\Delta \gamma^{N_{1}}_{S\Phi\ell_{j}}\,.
%\nonumber\\
%% --------- 3 ----------
%&= - \Delta\gamma^{N_{1}}_{S\Phi\ell_{i}}\sum_{j}{\cal  P}^{N_{1}}_{S\Phi\ell_{j}}
\end{align}
We have then used the specific relations valid for PFL 
(equivalent to $\sum_{j}\epsilon_{j}=0$)
\begin{align}
  \label{eq:asym-sum}
\sum_{j}\Delta\gamma^{N_{1}}_{S\Phi\ell_{j}}=
\sum_{j}\Delta\gamma^{N_{1}\bar{S}}_{\Phi\ell_{j}}=
\sum_{j}\Delta\gamma^{N_{1}\bar{\Phi}}_{S\ell_{j}}=
\sum_{j}\Delta\gamma^{\bar{\Phi}\bar{S}}_{N_{1}\ell_{j}}=0.
\end{align}
Thus, the last term in the second line of eq.~(\ref{eq:3to3-explanation}) vanishes, and 
the expression for $(\dot{Y}_{\Delta \ell_{i}})_{\Phi\ell_{i} S\leftrightarrow\Phi\ell_{j}S}$ 
given in (\ref{eq:3to3-complete}) is obtained.
%
% \begin{align}
%  \label{eq:3to3-final}
%  (\dot{Y}_{\Delta \ell_{i}})_{\Phi\ell_{i} S\leftrightarrow\Phi\ell_{j}S}
% = - \Delta\gamma^{N_{1}}_{S\Phi\ell_{i}}\sum_{j}{\cal
%   P}^{N_{1}}_{S\Phi\ell_{j}}\,.
% \end{align}
%
% Summing up these equations we finally obtain 
 %
% \begin{equation}
%   \label{eq:3to3-complete}
%   (\dot{Y}_{\Delta \ell_{i}})_{3\leftrightarrow 3}^{\text{sub}} =
%   - 2\left( {\cal P}^{N_{1}}_{S\ell_{i}\Phi}\Delta\gamma^{N_{1}}_{S\ell_{i}\Phi}
%   + ({\cal P}^{N_{1}\bar\Phi}_{S\ell_{i}}+{\cal P}^{N_{1}\bar S}_{\Phi\ell_{i}}
%   + {\cal P}^{\bar\Phi\bar S}_{N_{1}\ell_{i}})(\Delta\gamma^{N_{1}\bar\Phi}_{S\ell_{i}}
%   + \Delta\gamma^{N_{1}\bar S}_{\Phi\ell_{i}}
%   + \Delta\gamma^{\bar\Phi\bar S}_{N_{1}\ell_{i}})\right)\,.
% \end{equation}

 \subsection{$2\leftrightarrow 4$ processes}
\begin{figure}[t]
  \begin{center}
  \includegraphics[width=9.5cm,height=2.5cm,angle=0]{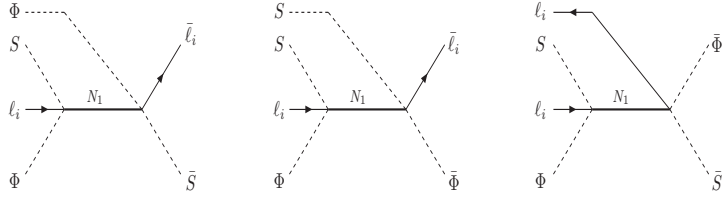}
  \end{center}
  \caption{Feynman diagrams for the $2 \leftrightarrow 4$ 
    with $|\Delta L_{i}|=2$ reactions $S\ell_{i}\leftrightarrow\bar\Phi\bar S
    \bar \ell_{i}\bar\Phi$, $\ell_i \Phi \leftrightarrow \bar S \bar S
    \bar\ell_{i}\bar\Phi$ and $\Phi S \leftrightarrow \bar\ell_{i}\bar S \bar
    \ell_{i}\bar \Phi $.}
  \label{fig:2to4}
  \end{figure}

  The inclusion of $2\leftrightarrow 4$ processes proceeds along 
 similar lines than for  $3\leftrightarrow 3$ processes.
As is shown by the diagrams in figure~\ref{fig:2to4}, there are three types  
of $|\Delta L_{i}|=2$ contributions to the evolution equation for  
$Y_{\Delta L_i}$. The on-shell pieces of these contributions can be
 written as:  
 \begin{equation}
   \label{eq:2to4reaction-densities}
   \gamma^{(\text{on})\,\ell_{i}S}_{\bar\Phi\bar S\bar\ell_{i}\bar\Phi} =
 P^{\ell_{i} S}_{\bar\Phi N_{1}}\gamma^{N_{1}}_{\bar S\bar\ell_{i}\bar\Phi},\qquad
   \gamma^{(\text{on})\,\ell_{i}\Phi}_{\bar S\bar S\bar\ell_{i}\bar\Phi} =
 P^{\ell_{i}\Phi}_{\bar S N_{1}}\gamma^{N_{1}}_{\bar S\bar\ell_{i}\bar\Phi},\qquad 
   \gamma^{(\text{on})\,\Phi S}_{\bar\ell_{i}\bar S\bar\ell_{i}\bar\Phi} =
   P^{\Phi S}_{\bar\ell_{i} N_{1}}\gamma^{N_{1}}_{\bar S\bar\ell_{i}\bar\Phi}\,.
 \end{equation}
By using again  eq.~(\ref{eq:off-shell-on-shel}) to relate the on-shell parts
to the relevant off-shell pieces, including the CP conjugate relations and   
including also the $2\leftrightarrow 4$ processes with 
$|\Delta L_{i}|=1$,  we obtain the following contributions:
 %
%  \begin{equation}
%    \label{eq:2to4contributions}
%    (\dot Y_{\Delta L_i})_{2\leftrightarrow 4}^{\text{sub}} =
%    (\dot Y_{\Delta L_i})_{\ell_i S\leftrightarrow \bar\Phi\bar S\bar\ell_i\bar\Phi}^{\text{sub}} +
%    (\dot Y_{\Delta L_i})_{\ell_i \Phi\leftrightarrow \bar S\bar S\bar\ell_i\bar\Phi}^{\text{sub}} +
%    (\dot Y_{\Delta L_i})_{\ell_i S\leftrightarrow \bar\ell_i\bar S\bar\ell_i\bar\Phi}^{\text{sub}}\,.
%  \end{equation}
 %
% By using again  eq.~(\ref{eq:off-shell-on-shel}) 
% The relevant on-shell parts of $2\leftrightarrow 4$ processes $|\Delta L_{i}|=2$ are
 %
 \begin{align}
   \label{eq:2to4-complete}
% --------- 1 ----------
   (\dot{Y}_{\Delta L_{i}})_{\Phi \ell_{i} S
     \Phi\leftrightarrow\bar{S}\bar{\ell}_{j}}
   &=-\Delta\gamma^{N_{1}\bar{\Phi}}_{S\ell_{i}} \sum_{j}{\cal
     P}^{N_{1}}_{S\Phi\ell_{j}}-\Delta\gamma^{N_{1}}_{S\Phi\ell_{i}}
   \sum_{j}{\cal P}^{N_{1}\bar{\Phi}}_{S\ell_{j}}\nonumber\\
% --------- 2 ----------
(\dot{Y}_{\Delta L_{i}})_{\Phi \ell_{i} S
  S\leftrightarrow\bar{\Phi}\bar{\ell}_{j}}
&=-\Delta\gamma^{N_{1}\bar{S}}_{\Phi\ell_{i}} \sum_{j}{\cal
  P}^{N_{1}}_{S\Phi\ell_{j}}-\Delta\gamma^{N_{1}}_{S\Phi\ell_{i}}
\sum_{j}{\cal P}^{N_{1}\bar{S}}_{\Phi\ell_{j}}\nonumber\\
% --------- 3 ----------
(\dot{Y}_{\Delta L_{i}})_{\Phi \ell_{i} S
  \ell_{j}\leftrightarrow\bar{S}\bar{\Phi}}
&=-\Delta\gamma^{\bar{\Phi}\bar{S}}_{N_{1}\ell_{i}} \sum_{j}{\cal
  P}^{N_{1}}_{S\Phi\ell_{j}}-\Delta\gamma^{N_{1}}_{S\Phi\ell_{i}}
\sum_{j}{\cal P}^{N_{1}\bar{\ell}_{j}}_{\Phi S}\nonumber\\
% --------- 4 ----------
(\dot{Y}_{\Delta L_{i}})_{\Phi \ell_{i} S \bar{\Phi}\leftrightarrow S\ell_{j}}
&=-\Delta\gamma^{N_{1}}_{S\Phi\ell_{i}} \sum_{j}{\cal
  P}^{N_{1}\bar{\Phi}}_{S\ell_{j}}-\Delta\gamma^{N_{1}\bar{\Phi}}_{S\ell_{i}}
\sum_{j}{\cal P}^{N_{1}}_{S\Phi\ell_{j}}\nonumber\\
% --------- 5 ----------
(\dot{Y}_{\Delta L_{i}})_{\Phi \ell_{i} S \bar{S}\leftrightarrow\Phi\ell_{j}}
&=-\Delta\gamma^{N_{1}}_{S\Phi\ell_{i}} \sum_{j}{\cal
  P}^{N_{1}\bar{S}}_{\Phi\ell_{j}}- \Delta\gamma^{N_{1}\bar{S}}_{\Phi\ell_{i}}
\sum_{j}{\cal P}^{N_{1}}_{S\Phi\ell_{j}}\nonumber\\
% --------- 6 ----------
(\dot{Y}_{\Delta L_{i}})_{\Phi \ell_{i} S \bar{\ell}_{j}\leftrightarrow S
  \Phi} &=-\Delta\gamma^{N_{1}}_{S\Phi\ell_{i}}\sum_{j}{\cal
  P}^{N_{1}\bar{\ell}_{j}}_{\Phi S}-\Delta\gamma^{\bar{\Phi}
  \bar{S}}_{N_{1}\ell_{i}}\sum_{j}{\cal P}^{N_{1}}_{S\Phi\ell_{j}}.
 \end{align}

  From the sets of equations~(\ref{eq:1to3and2to2-complete}),
  (\ref{eq:3to3-complete}) and (\ref{eq:2to4-complete}), and using
  $\sum_{j}({\cal P}^{N_{1}}_{\Phi\ell_{j} S} + {\cal P}^{N_{1} \bar
    S}_{\ell_{j}\Phi} + {\cal P}^{N_{1}\bar\Phi}_{\ell_{j} S} + {\cal
    P}^{N_{1}\bar\ell_{j}}_{S\Phi}) = 1$, we finally obtain the BE for the
  evolution of $Y_{\Delta L_i}$:
 \begin{align}
   \label{eq:flavored-asymm-EvEq}
   \dot{Y}_{\Delta L_{i}} &= 
   (y_{N_{1}}-1)(
   \Delta\gamma^{N_{1}}_{S\ell_{i}\Phi}
   +\Delta\gamma^{N_{1}\bar{S}}_{\Phi\ell_{i}}+\Delta\gamma^{N_{1}\bar{\Phi}}_{S\ell_{i}}
   +\Delta\gamma_{S\Phi}^{N_{1}\bar\ell_{i}})\\
   &  - \Delta y_{\ell_{i}}(
   \gamma^{N_{1}}_{S\ell_{i}\Phi}+\gamma^{N_{1}\bar{S}}_{\Phi\ell_{i}}
   + \gamma^{N_{1}\bar{\Phi}}_{S\ell_{i}}
   + y_{N_{1}}\gamma_{S\Phi}^{N_{1}\bar\ell_{i}})\nonumber\\
   & - \Delta y_{\Phi}(
   \gamma^{N_{1}}_{S\ell_{i}\Phi}+\gamma^{N_{1}\bar{S}}_{\Phi\ell_{i}}
   + y_{N_{1}}\gamma^{N_{1}\bar{\Phi}}_{S\ell_{i}}
   + \gamma_{S\Phi}^{N_{1}\bar\ell_{i}})\nonumber\\
   & - \Delta y_{S}(
   \gamma^{N_{1}}_{S\ell_{i}\Phi}+y_{N_{1}}\gamma^{N_{1}\bar{S}}_{\Phi\ell_{i}}
   +\gamma^{N_{1}\bar{\Phi}}_{S\ell_{i}}
   +\gamma_{S\Phi}^{N_{1}\bar\ell_{i}})\nonumber\,.
 \end{align}
where the  term $(y_{N_{1}}-1)$ in the r.h.s. shows that the correct
thermodynamical behavior is recovered.

Finally, using the approximate equalities between the scatterings and the decay
 asymmetries~ \cite{Nardi:2007jp}
 \begin{equation}
   \label{eq:decay-scattering-eq}
   \frac{\Delta \gamma^{N_1}_{\ell_i\Phi S}}{\gamma^{N_1}_{\ell_i\Phi S}}\simeq
   \frac{\Delta \gamma^{N_1 \bar\Phi}_{\ell_i S}}{\gamma^{N_1 \bar\Phi}_{\ell_i S}}\simeq
   \frac{\Delta \gamma^{N_1 \bar S}_{\ell_i\Phi}}{\gamma^{N_1 \bar S}_{\ell_i\Phi}}\simeq
   \frac{\Delta \gamma^{N_1 \bar\ell_i}_{\Phi S}}{\gamma^{N_1\bar\ell_i}_{\Phi S}}\,,
 \end{equation}
the source term in equation (\ref{eq:flavored-asymm-EvEq}) can be rewritten as
 \begin{equation}
   \label{eq:boltzmann-eq-almostFinal}
   \left(\dot{Y}_{\Delta L_{i}}\right)_{\rm source} = (y_{N_1} - 1) 
   \frac{\Delta \gamma^{N_1}_{\ell_i\Phi S}}{\gamma^{N_1}_{\ell_i\Phi S}}\>   \gamma_i 
% - \Delta y_{\ell_i}   \left[
%     \gamma_i + (y_{N_1} - 1)\gamma^{N_1\bar\ell_i}_{S\Phi}   \right]   
\,,
 \end{equation}
 where $\gamma_i = \gamma^{N_1}_{\ell_i\Phi S} + \gamma^{N_1 \bar\Phi}_{\ell_i S}
 + \gamma^{N_1 \bar S}_{\ell_i\Phi} + \gamma^{N_1 \bar\ell_i}_{\Phi S}$. By using the relation
 \begin{equation}
   \label{eq:decay-scattering-relation}
   \frac{\gamma^{N_1}_{\ell_i \Phi S}}{\sum_{j}(\gamma^{N_1}_{\ell_j \Phi S} + 
     \gamma^{N_1}_{\bar\ell_i \bar\Phi \bar S})} =
   \frac{\gamma_i}{\sum_j(\gamma_j +
 \bar\gamma_j)}
%{\gamma_{\text{tot}}}
\,,
 \end{equation}
% 
% where $\gamma_{\text{tot}}=\sum_i(\gamma_i + \bar\gamma_i)=2\sum_i\gamma_i$, 
and recalling that the flavored CPV asymmetries are given by
 \begin{equation}
   \label{eq:flavored-CPV-symmetries-app}
   \epsilon_i = \frac{\Delta\gamma^{N_1}_{\ell_i\Phi S}}
   {\sum_j(\gamma^{N_1}_{\ell_j\Phi S} + \gamma^{N_1}_{\bar\ell_j\bar\Phi\bar S})}
 \end{equation}
 the evolution equation in~(\ref{eq:boltzmann-eq-almostFinal}) can be recast as
%
% \begin{equation}
\begin{align}
   \label{eq:boltzmann-eq-final}
     \dot{Y}_{\Delta L_{i}} = & \epsilon_{i} 
   \left(y_{N_1} - 1 \right)\gamma_{\text{tot}} 
- \Delta y_{i}\left[\gamma_{i} + \left(y_{N_1} -
   1\right)\gamma^{N_{1}\bar\ell_{i}}_{S\Phi}\right] \nonumber \\
&- \Delta y_{\Phi}\left[\gamma_{i} + \left(y_{N_1} -
   1\right)\gamma^{N_{1}\bar\Phi}_{S\ell_{i}}\right]  
% \nonumber \\ &
- \Delta y_{S}\left[\gamma_{i} + \left(y_{N_1} - 1\right)\gamma^{N_{1}\bar
   S}_
{\Phi\ell_{i}}\right]
\,, 
\end{align}  
% \end{equation}       
%
where $\gamma_{\text{tot}}$ is given in eq.~(\ref{eq:gammatot}).  Neglecting
the terms in the second line proportional to $\Delta y_{\Phi}$ and $\Delta
y_{S}$ amounts to neglect ${\cal O}(1)$ effects~\cite{Nardi:2005hs}, and still
yields a quite reasonable approximation. Finally, the evolution equation for
the heavy Majorana neutrino can be simply written as:
 \begin{equation}
   \label{eq:right-handed-Neutrino-DENS}
   \dot Y_{N_1} = -(y_{N_1} - 1)\,\gamma_{\text{tot}}\,.
 \end{equation}

 \section{Decay and Scatterings}
 \label{sec:appendixB}
 The thermally averaged reaction density for $N_1\leftrightarrow
 \ell_i\Phi S$ is given by~\cite{Luty:1992un,Giudice:2003jh}
 \begin{equation}
   \label{eq:decay-reaction-dens}
   \gamma^{N_1}_{\ell_i\Phi S}= N_{N_1}^{\text{eq}}\,\frac{K_1(z)}{K_2(z)}
   \,\Gamma^{N_1}_{\ell_\Phi S}\,,
 \end{equation}
 where $N_{N_1}^{\text{eq}}$ is the  equilibrium
 number density for $N_1$, $K_{1,2}(z)$ are Bessel functions (see appendix B in
 ref.~\cite{Davidson:2008bu}) and $\Gamma^{N_1}_{\ell_\Phi S}$ is the
 decay width given in (\ref{eq:total-decay-width}). 
 The  thermally averaged $2\to 2$ reaction densities are given by 
 \begin{equation}
   \label{eq:general-gamma-scatt}
   \gamma_{2\to 2} = \frac{M^{4}_{N_1}}{512\pi^{5}z}\int^{\infty}_{1}dx\sqrt{x}K_{1}
   \left(z\sqrt{x}\right)\hat{\sigma}\left(x\right)\,.
 \end{equation}
 Here $\hat\sigma(x)$ is the dimensionless reduced cross section 
 \begin{equation}
   \hat{\sigma}%_{\text{chan}}
 (x) =
   \sum_{a} r^{2}_{a} h^{*}_{ia} h_{ia}
   \lambda^{*}_{1a}\lambda_{1a}F^{a}
 % _{\text{chan}}
 (x)
   + 2\sum_{a<b}r_{a}r_{b}\,\mathbb{R}\text{e}[h^{*}_{ia}h_{ib}\lambda^{*}_{1a}
   \lambda_{1b} G^{a,b}
 % _{\text{chan}}
 (x)].
 \end{equation}
 The explicit form of the kinematical functions $F^{a}%_{\text{chan}}
 (x)$ and
 $G^{a,b}
 %_{\text{chan}}
 (x)$ depends on the specific   $s$, $t$ or $u$ channel processes.
 In practice, for the $t$ and $u$ channels it is always a good approximation to
 use  point-like interactions  obtained by integrating out 
 the heavy vectorlike fields. In this approximation,  
 $\gamma^{N_1\bar\Phi}_{S\ell_i}
 =\gamma^{N_1\bar\ell_i}_{\Phi S}=\gamma^{t,u}_{2\to 2}$ where 
 \begin{equation}
   \label{eq:point-like-reaction-dens}
   \gamma_{2\to 2}^{t,u} = \frac{M^{4}_{N_1}}{1024\pi^{5}z}
   \left[
     \sum_{a} r^{2}_{a} h^*_{ia} h_{ia}
   \lambda^*_{1a}\lambda_{1a}
   + 2\sum_{a<b}r_ar_b\,\mathbb{R}\text{e}(h^*_{ia}h_{ib}\lambda^*_{1a}\lambda_{1b})
   \right]\,f(z), 
 \end{equation}
 with
 \begin{equation}
   \label{eq:fz}
   f(z)=\int_1^\infty\,dx\,\frac{x^2 - 1}{\sqrt{x}}\,K_1(z\sqrt{x})\,.
 \end{equation}
 For $s$-channel scatterings ($N_1 \bar S\leftrightarrow \ell_i \Phi$) the
 pointlike approximation is not sufficiently accurate, expecially at high
 temperatures $T> M_{N_1}$, and the complete expression for the kinematical
 functions has to be used. We define
 \begin{equation}
   \label{eq:s-channel-kin-func2}
   G^{a,b}_{s}(x)= \frac{x - 1}{H^{a,b}_{s,1}}\times[(1-x)\,H^{a,b}_{s,2} 
   + (1+x)\,H^{a,b}_{s,3}]\,,
 \end{equation}
 with  
 \begin{align}
   \label{eq:s-channel-newfun1}
   H^{a,b}_{s,1}(x) &= 2 x (1-r_{a}^{2}x-2i r_{a}\eta_{a})
   (1-r_{b}^{2}x+2i r_{b}\eta_{b}),\\
   \label{eq:s-channel-newfun2}
   H^{a,b}_{s,2}(x) &= 2r_{a}r_{b}+r_{a}+r_{b} + ir_{a}r_{b}(\eta_{a} 
   - \eta_{b}) + r_{a}r_{b}(x - 1),\\
   \label{eq:s-channel-newfun3}
   H^{a,b}_{s,3}(x) &= r_{a}r_{b}+r_{a}+r_{b}+ir_{a}r_{b}(\eta_{a}-\eta_{b})
   + (1 + ir_{a}\eta_{a})(1-ir_{b}\eta_{b})\,,   
 \end{align}
and  
 \begin{equation}
   \label{eq:s-channel-kin-func1}
   F^{a}_{s}(x) = G^{a,a}_{s}(x)\,.
 \end{equation}

 The dimensionless parameter $\eta_a$ introduced in the equations above
 corresponds to the total decay widths of the messenger fields $F_a$ normalized
 to the $N_1$ neutrino mass:
 \begin{equation}
  \eta_{a}=\frac{\Gamma_{a}}{M_{N_1}}=\frac{1}{8\pi r_{a}}\left[ 
  (1-r^{2}_{a})(1+r_{a})^{2}\lambda^{*}_{1a}
  \lambda_{1a}+\frac{1}{2}\sum_{i}h^{*}_{ia}h_{ia}\right]\,.
  \end{equation}

 \end{document}